\documentclass[a4paper,10pt]{article}
\usepackage[geometry]{ifsym}
\usepackage[english]{babel}
\usepackage{verbatim}
\usepackage{natbib}
\usepackage{html}
\usepackage{color}
\usepackage{soul}
\usepackage{mathrsfs}
\usepackage{amsfonts,amsmath,latexsym}
\usepackage[T1]{fontenc}
\usepackage[utf8x]{inputenc}
\usepackage{graphicx}
\usepackage{tabularx,ragged2e,booktabs}
\usepackage[font=small,labelfont=bf,labelsep=period,figurename={Fig.}]{caption}
\usepackage{subcaption}
\usepackage{epstopdf}
\usepackage{comment}
 \usepackage{enumerate}
 \usepackage{epsf}
 \usepackage{amsthm}
 \usepackage{color}
\usepackage{amssymb}
\usepackage{fancyhdr} 
\usepackage{palatino}
\newtheorem{thm}{Theorem}[section]

\newtheorem{lemma}[thm]{Lemma}
\newtheorem{coro}[thm]{Corollary}
\newtheorem{define}[thm]{Definition}
\newtheorem{rmk}[thm]{Remark}

\newtheorem{dem}{Proof}

\def\R{\mathbb R}
\def\N{\mathbb N}
\def\Z{\mathbb Z}

\author{
{\sc Dario Gasbarra}
\thanks{Corresponding author, Department of Mathematics and Statistics,  University of Helsinki   P.O. Box 68 
FI-00014   Finland e-mail:{ \tt dario.gasbarra@helsinki.fi}} 
 {\sc,}\ {\sc Jia Liu}
\thanks{  Corresponding author,     Department of Mathematics and Statistics,  University of Jyv\"askyl\"a, P.O.Box (MaD) FI-40014  Finland e-mail:{\tt  jia.liu@jyu.fi}}
\ {\sc and}\ {\sc Juha Railavo} 
\thanks{HUS e-mail:{\tt juha.railavo@elisanet.fi}}.
}

\date{\today}
\title{  
Data augmentation in  Rician noise model and Bayesian Diffusion Tensor Imaging. }
\newcommand{\MBFigure}[6]{
$\left. \right.$ \\
\refstepcounter{figure}
\addcontentsline{lof}{figure}{\numberline{\thefigure}{\ignorespaces #5}}
\begin{center}
\begin{minipage}{#1cm}
\centerline{\includegraphics[width=#2cm,angle=#3]{#4}}
\begin{center}
\upshape{F\textsc{ig} \normal
\end{center}
size{\thefigure}. $-$} #5
\end{center}
\label{#6}
\end{minipage}
\end{center}
$\left. \right.$ \\}

\begin{document}
\maketitle 
\begin{abstract} 

Mapping white matter tracts is  an essential step towards  understanding brain function.
Diffusion Magnetic Resonance Imaging (dMRI)
is the only noninvasive technique which can detect in vivo anisotropies in the 3-dimensional diffusion of water molecules,
which correspond to nervous fibers in the living brain. In this process,
spectral data from the displacement distribution of water molecules is collected by a magnetic resonance scanner.
From the statistical point of view,  inverting the Fourier transform from such sparse and noisy spectral measurements
leads to a non-linear regression problem.
Diffusion tensor imaging (DTI) is the simplest  modeling approach postulating a 
 Gaussian displacement distribution at  each volume element (voxel). 
 Typically the inference is based on a linearized log-normal regression model
that can fit the spectral data at low frequencies. However such approximation fails to fit
  the high frequency measurements which 
contain information about the details of 
the displacement distribution but have a low signal to noise ratio. 
In this paper, we directly work with the Rice noise model and cover the full range of $b$-values.
 Using data augmentation to represent the likelihood, we reduce
the non-linear regression problem to the framework of generalized linear models.
Then we construct a Bayesian hierarchical model in order to  perform simultaneously estimation and regularization
of the tensor field. Finally the Bayesian paradigm is implemented by  using Markov chain Monte Carlo.

\if
This work is motivated by the need of diagnostics for the Lewy 
bodies disease.
\fi


\end{abstract}



\section{ Introduction}


Diffusion as a physical phenomenon has been an essential part of the history and development of magnetic
 resonance imaging.
\cite{hahn} observed the effect of diffusion to spin-echoes,
 \cite{carr} 
 studied the effects of diffusion on free precession,
 and \cite{torrey}  modified the
 Bloch equations to include diffusion term with spatially varying magnetic field.
\cite{stejskal}, 
in their seminal paper, introduced the pulsed gradient spin echo sequence
 and showed the potential of diffusion related signal attenuation to probe the motion of molecules
 and to define the diffusion coefficient.
In 1973  P. Lauterbur (who shared the
 Nobel Prize with Sir Peter Mansfield in 2003) made history  publishing  his groundbreaking paper entitled
``Image formation by induced local interactions: Examples employing
 nuclear magnetic resonance'' 
 In his experiment Lauterbur superimposed a magnetic field gradient on the static uniform magnetic field.
 Because of the Larmor principle, different parts of the sample would have different resonance frequencies and so a given 
resonance frequency could be associated with a given position. He also pointed out that
it is possible to measure  molecular diffusion from the decay of the MR-signal.
 Diffusion weighted magnetic resonance imaging was introduced by \cite{lebihan86} measuring the displacement of protons.
 \cite{moseley} observed that diffusion in the white matter was anisotropic. 
In anisotropic media  the mobility of the molecules is orientation dependent and can not
 be represented by one single diffusion coefficient. The three dimensional process of diffusion 
 modeled by diffusion tensors was introduced  by \cite{basser-94}. 

Without going into the physics of dMRI, we sketch 
the idea from the statistical point of view.
After applying two consecutive and opposite gradient pulses with amplitude   $|{\bf q}|$
in the direction  ${\bf u}={\bf q}/|{\bf q}| \in S^2$,\footnote{ $S^2\subset \R^3$ denotes the unit sphere.} with time
delay $t$, MR produces at every spatial location  $v$ a signal
\begin{align} 
\label{eq:signal1} 
 S_v({\bf q})  \; = \;
S_v({\bf 0 }) E_{v}\biggl( \exp\bigl( i \;{\bf q} \cdot {\bf V}_t  \bigr ) \biggr) \;
= \; S_v({\bf 0 })  \exp\biggl(  -  \frac 1 2 \; 
 {\bf q}  {D}_v  {\bf q^{\top} }\biggr) = S_{v} ({\bf  0} )   \exp\biggl(  -  b  
 {\bf u}  {D}_v  {\bf u^{\top} }\biggr)
\end{align}
 where $S_v({\bf 0})$ is the concentration of water molecules at $v$, and ${\bf q}$ is the 3-dimensional pulse gradient,
$b=|{\bf q}|^2/2$.
\nolinebreak
In
eq. (\ref{eq:signal1})   appears
the characteristic function of a centered Gaussian random vector ${\bf V}_t$ with 
covariance matrix ${D}_v$\footnote{In the neuroimaging literature another convention is used, with
$D= E\bigl({\bf V}_t^{\top} {\bf V}_t\bigr)/2$  and $b=|{\bf q}|^2$.},
which is interpreted as  the displacement of a 
water molecule with initial position $v$ in the time interval $[0,t]$ between the two pulses.
The symmetric and positive definite matrix-valued field $({D}_v)$  describes the geometry of the media
and it is  the object of interest.
Note that for an eigenvector ${\bf q}$ with eigenvalue $\lambda > 0$ satisfying
 $D_v {\bf q}=  \lambda {\bf q}$, the MR signal
\begin{align} 
\label{eq:Zeig} 
  S_v( {\bf q} )=S_v({\bf 0}) \exp\bigl( - \frac  12 \lambda |{\bf q}|^2  )
\end{align}
is highest when ${\bf q}$ belongs to the eigenspace of  the smallest
eigenvalue of $D_v$, and lowest in the  principal direction.
 In neuroimaging,  we measure restricted diffusion within neuron cells, and
 the principal diffusion eigenvector corresponds to the direction of a nervous fiber.

It is well known that the noise in an MR measurement has a Rice distribution
 instead of Gaussian \citep{jones,henkelman,ibrahim,asselmal,landman}. Several 
 authors \citep[e.g.][]{ibrahim, salvador}, 
  add the noise-induced bias
 into the measurement so that a simple Gaussian noise model can be fitted to the data.
But none of them can easily gain the potential important information 
\citep[e.g.][]{Mori,burdette} from the high-frequency data,
 because in the high $b$-value range the corrected data does
 not fit the Gaussian distribution. Also
 the Rice noise model is used \citep[e.g.][]{gudbjartsson,Jelle,andersson2008,Lauwers},
 but in all cases the methods dealing with Rice noise are computationally intensive. 

Our work also deals  directly with the Rice noise distribution. 
By using data augmentation, we reduce  the non-standard  regression
 problem to a standard Poisson regression.
This novel strategy can obtain diffusion information also from  
high amplitudes in the low SNR regime, including  the zero  measurements
which fall below the detection threshold.
Bayesian regularization is introduced  in order to reduce 
the noise  and obtain   estimates also  when the data is locally corrupted  and contains  artefacts.
\nolinebreak
In addition, our method  applies directly to high-order tensor models and spherical harmonics
expansions of the diffusivity function
see \cite{barmpoutis}, \cite{ozars}, \cite{ghosh},
 which can capture more complex brain structures as fiber crossings and branchings.
In order to regularize the 4th order tensor field
we use a recent result by \cite{GPD-inria} on invariants of 4th order tensors
to derive the  general
form of an isotropic  Gaussian distribution
for the tensor coefficients.
 This generalizes the probabilistic models 
 proposed in the literature \citep[see][]{pajevic, moakher}.

The paper is structured as follows: 
the nonlinear regression problem with Rician noise model is described
in  
  Section \ref{Rice:section}. 
The main contribution of  the paper, 
data-augmentation by Poissonization is introduced in Section \ref{data:augmentation:section}. 
In Section \ref{bayes:modeling}, after a general discussion on McMC methods,
we construct the Bayesian hierarchical model for a single tensor (Section \ref{sub:hierarchical}),
and the Gibbs-Metropolis algorithm  
for sampling  posterior distribution (Section \ref{mcmc:update}).  
In Sections \ref{dependence_prior} and \ref{regularization},
we continue with the isotropic Gaussian  Markov field prior 
the Gibbs-Metropolis updates for the  tensor field together with the
Bayesian estimation of the regularization parameters.
In Sections   \ref{sub:4thorder_model},\ref{sub:update:regularization_4},
\ref{SH:subsection} we extend the method to higher order tensor models and explain the correspondences between tensors and the 
spherical harmonic expansion of the diffusivity.
The implementation of these methods is illustrated
in  Section \ref{results}
 with an analysis of human brain data.

\section{Theory and Modeling}
  
\subsection{ Rice likelihood} \label{Rice:section}

We follow the discussion in  \cite{ibrahim}.
Let us fix a position $v$ and omit the indexing. The signal is 
expressed conveniently as $S_v({\bf q})= \exp( Z \theta)$
with parameter
\begin{align*} 
\theta=(\theta_0,\theta_1,\dots, \theta_d)^{\top}:=  
\bigl(  \log S({\bf 0}) ,
 D_{xx}, D_{yy} , D_{zz}, D_{xy}, D_{xz}, D_{yz} \bigr)^{\top}
\end{align*}
and the design matrix $Z$ has rows
\begin{align*}
Z({\bf q}) =  \bigl(  1, -{\bf q}_x^2/2 ,   -{\bf q}_y^2/2, - {\bf q}_z^2/2,  
-{\bf q}_x  {\bf q}_y , -{\bf q}_x  {\bf q}_z , -{\bf q}_y  {\bf q}_z  \bigr)\; . 
\end{align*} 
In the MR experiment the signal is corrupted by Rice noise.  We measure
\begin{align*} 
   Y({\bf q})=  \big \vert S_v({\bf q}) + \varepsilon \big\vert = \sqrt{ \bigl 
( \exp( Z\theta) + \varepsilon_1 \bigr)^2 + \varepsilon_2^2 },
\end{align*}
where  $(\varepsilon_1, \varepsilon_2)$ are independent with Gaussian distribution ${\mathcal N}(0,\sigma^2)$,
and  $\varepsilon= 
( \varepsilon_1 + i \varepsilon_2)$ is a complex Gaussian noise.

From the statistical point of view,
the estimation of $\theta$ from diffusion-MR data is a non-linear regression problem 
with the positivity constraint 
$\bigl({\bf q} D {\bf q}^{\top} \bigr) \ge 0 , \;$ for all ${\bf q}\in \R^3$.
It follows that the Rice likelihood function is given by
\begin{align} \label{rice:density} 
& p_ {\theta, \sigma^2}( y | Z ) = \frac{ y }{\sigma^2} 
\exp\biggl(  - \frac{ y^2+\exp( 2Z\theta) }{2\sigma^2} \biggr)
I_0\biggl(  \frac{ y\exp( Z\theta) }{\sigma^2}\biggr) ,
\end{align}
where 
\begin{align}
  I_0(z)=  \frac 1 {\pi} \int\limits_0^{\pi} \exp(z \cos t )dt \end{align}
is the modified Bessel function of first kind. 

Diffusion-MR data $(Y_i,Z_i)$ is collected for a series of pulses $({ \bf q}_i:i=1,\dots,m)\subset\R^3$.
Direct maximum likelihood estimation of the parameters $(\theta, \sigma^2)$  from the sampling 
density of Eq. $(\ref{rice:density})$ is problematic, involving the numerical evaluation
of modified Bessel functions. A simplified  popular approach is to approximate the Rice likelihood of Eq. $(\ref{rice:density})$ by a log-normal
model for $Y$, where $\log(Y)$ is Gaussian with mean $(Z\theta)$ and variance   $\sigma^2\exp( -2Z\theta)$.
The model parameters are then estimated by using  iterated Weighted Least Squares (WLS) \citep[see][]{ibrahim,koay}.
 However this approximation works well only within a certain narrow range of amplitudes.
 In clinical studies and  research papers, 
 most often the maximal 
 $b$-value is in the range of    $ 600 -1200 s/mm^2$ \citep{Mori},\cite{jones,ibrahim,hagmann,peter}.
Within this range the log-normal approximation to the Rice noise is adequate. However, for large  $b$-values,
 the SNR is low,  the data  does not fit 
 the log-normal approximation, and the WLS-algorithm may fail to converge. Reports \citep[e.g.][]{gudbjartsson}
 address more than half underestimation of the true noise based the Gaussian model. Moreover, since the data is digitalized, 
at high $b$-values one may get measurements $Y_i$ which are
coded as zeros. In order to use the log-normal approximation, 
these zero values have to be discarded, inducing sampling bias. 
When the estimation concerns only of 2nd-order tensors, under the assumption of
 Gaussian diffusion,  it  is enough to use low $b$-value measurements.
However,  to estimate higher order characteristics and finer details
of the diffusion distribution using higher order tensor models, expansions of spherical functions
or mixture models,  and ideally,
 to invert the characteristic function in Eq. (\ref{eq:signal1}) in the non-Gaussian case, the high $b$-value measurements are
 also needed. 

\subsection{Poissonization and data augmentation }\label{data:augmentation:section}

From  a  statistician's point of view, a non-linear regression problem is most conveniently
 framed in the context of  {\it Generalized Linear Models} (GLM), where the measurements have
probability density of the form
\begin{align}\label{eq:glm} 
p_{\theta,\phi }(y | Z)= f_{\tau,\phi}(y)=c(y,\phi) \exp\biggl(\frac{ y\tau- a(\tau) }{\phi} \biggr) \;, 
\end{align}
see \cite{mc_cullagh-nelder}.
The function $a(\tau)$ in Eq. (\ref{eq:glm}) specifies an exponential family of distributions for the response $Y$, and 
 $\tau$ is determined implicitly by the relation $g(\mu)=Z\theta$,
where  $\mu= E_{\tau,\phi}(Y)=a'(\tau)$ is the expectation and $g(\mu)$ is the {\it link} function.
Unfortunately, this assumption is  not satisfied by the Rice  likelihood in Eq. (\ref{rice:density}).
In order to reduce the  non-linear regression problem to the framework of generalized linear models, 
 we  propose a novel data augmentation strategy  for  parameter estimation under
the  exact Rice likelihood.
For each data point $Y$ we introduce an unobservable
variable  $N$ which follows a generalized linear model with Poisson response corresponding to $a(\tau)=\exp(\tau)$, $\phi=1$,
and link function $g(\mu)=\log(2 \sigma^2 \mu)/2$. In a Bayesian framework, we then use Markov chain Monte Carlo
to  integrate, conditionally on the observations $Y$, the variables   $\theta, \sigma^2$ and $N$.

\begin{lemma} \label{representation:lemma}Consider random variables $(N,X)$, where
 $N$ is Poisson distributed with mean $t>0$ , and given $N$, $X$ has conditional distribution $\mbox{Gamma}( N+1, 1/ (2\sigma^2) )$,
that is 
\begin{align*} 
  P_{t,\sigma^2} (N=n, X \in dx) = P_t( N=n) P_{\sigma^2}(X\in dx| N=n)=
  \frac {(tx) ^{n }   }{ (n!)^2  (2\sigma^2)^{n+1} } \exp\biggl(-t -  \frac x {2\sigma^2} \biggr ) dx \; .
\end{align*}
Then
\begin{enumerate}
 \item \label{wikipedia}
 $Y:=\sqrt{ X} $ has marginal density
\begin{align*} 
P_{t,\sigma^2}( Y \in dy) = 
\frac y {\sigma^2} \exp\biggl( -t - \frac{ y^2}{ 2\sigma^2} \biggr) I_0 \biggl( \frac{ y }{\sigma }\sqrt{ 2t  } \biggr) dy
\end{align*}
\item
 The conditional distribution of $N$ given $Y$ is
\begin{align} \label{N|Y}
  P_{t,\sigma^2} ( N=n | Y=y )=    I_0\biggl( \frac{ y }{\sigma} \sqrt{2t} \biggr)^{-1}
\biggl(
 \frac{   y^2 t  }{2\sigma^2} \biggr)^{n} (n!)^{-2 }
\end{align}
In particular  $P_{t,\sigma^2}(N=0 | Y=0)=1$.
\end{enumerate}
\end{lemma}
\begin{dem} \ref{wikipedia} is well known.
After a change of variable sum  over $n$ by using the representation 
\begin{align}\label{special_representation} 
I_0(2z)= {}_0 F_1(1,z^2)= \sum_{n=0}^{\infty}  \frac{ z^{2n} } {(n!)^2}
\end{align} 
\citep{tables}, 
where ${}_0 F_1(1,z)$ is a Gaussian hypergeometric function.
Eq. (\ref{N|Y}) is a consequence of the Bayes formula.
\end{dem}

\begin{define} For $\tau>0$, consider two i.i.d. random variables $N, N'$ with Poisson($\tau$) distribution, and define the
probability distribution
\begin{align*}
       p_{\tau}( n ):= 
P_{\tau}( N=n  | N=N') = I_0(2\tau)^{-1} 
\frac{ \tau^{2n} }{(n!)^2 } ,\quad  n \in \N.
\end{align*}
We call $(p_{\tau}(n) :n\in \N)$ the {\it reinforced Poisson distribution } with parameter $\tau$.
\end{define}
In  appendix (\ref{reiforced:sampling}) we discuss random sampling from this distribution.

\begin{coro} \label{representation:coro}
 In the settings of Lemma \ref{representation:lemma}, with $t=\exp( 2 Z\theta ) / (2\sigma^2 )$ ,
\begin{itemize}\item 
The marginal distribution of 
$Y$ has Rice density of Eq. $(\ref{rice:density})$.
\item The conditional distribution $P_t( N = n |  Y=y)$ is  a reinforced Poisson distribution $p_{\tau}(n)$ with parameter 
\begin{align*}
\tau=\frac{ y\exp(Z\theta) }{ 2\sigma^2}  \;.
\end{align*}
\end{itemize}
\end{coro}


\section{Bayesian Computational Methods } \label{bayes:modeling}

\subsection{Markov chain Monte Carlo}\label{sec:4} 

The Metropolis-Hastings algorithm \citep{metropolis, hastings} 
is  a general method to explore a probability distribution  in 
high-dimensional space. The idea  is to construct 
a Markov chain $(\xi_t)$ which is reversible with respect to
the target probability $\pi(x)$, i.e. the transition probability  $K(x\to dy)=P(\xi_1 \in dy | \xi_0 =x)$
satisfies the {\it detailed balance condition}
\begin{align*}
 P_{\pi }( \xi_0  \in dx,   \xi_1  \in dy )=   \pi(dx)  K( x \to  dy)  =
\pi(dy) K(y\to dx) = P_{\pi}( \xi_0  \in dy, \xi_1 \in dx) \; .
\end{align*}
It follows that  $\pi$ is the {\it  equilibrium distribution}  of the Markov chain, meaning that 
the Markov chain starting from the equilibrium distribution remains in equilibrium, i.e.
 $\pi(dx)=P_{\pi}( X_t \in dx ), \; \forall t\in \N$.

  Let $\pi(x)=z^{-1} f(x)$ be the target probability density for a
 configuration $x \in \R^d$,
 where 
\begin{align*}
z=\int\limits_{\R^d} f(x) dx <\infty
\end{align*}  is a possibly unknown normalizing
constant. Starting from  a configuration $\xi_{t}$,
sample a proposal value $\widetilde \xi$
from a proposal density $Q( \xi_{t}\to \widetilde  \xi)$.
 With  probability
\begin{align} \label{MHacceptance}  A( \xi_t \to \widetilde \xi):=
\min\biggl\{   \frac{ f(\widetilde \xi) Q( \widetilde \xi \to \xi_t) }
{ f(\xi_t) 
Q(\xi_t\to\widetilde\xi)}
 , 1 \biggr\} ,
\end{align} we
accept the proposed value and set $\xi_{t+1}=\widetilde \xi$, otherwise
the proposed move is rejected and we set $\xi_{t+1}=\xi_t$.
The ratio of densities in the right hand side of Eq. (\ref{MHacceptance}) is referred as Hastings' ratio.
 It is straightforward to
check that the  resulting transition probability 
\begin{align*}
P( \xi_{t+1}\in d\xi | \xi_t)=
  K( \xi_t\to  d\xi ) =  A( \xi_t \to \xi)  Q(  \xi_t \to d\xi )+ \delta_{\xi_t}( d\xi ) \int_{\R^d} \bigl (1-A(\xi_t \to \eta)\bigr) Q( \xi_t \to d\eta) 
\end{align*}
satisfies detailed balance  and the Markov chain $(\xi_t)$
is reversible    with respect to the target distribution $\pi$.                                                                
 In order to implement
the algorithm, it is enough to know the target density up to a proportionality constant.


  Note that  when we apply consecutively different
Metropolis-Hastings transitions, the equilibrium distribution is preserved.
 Under some irreducibility assumptions,
the Markov chain covers the support of the target distribution
\citep[see][]{nummelin}, and  the  ergodic theorem 
\begin{align*}
  \lim_{T\to\infty}  \frac 1  T \sum_{t=0}^{T-1} g(\xi_t) = \frac{ \int_{\R^d} g(x) f(x) d x }
{ \int_{\R^d}  f(x) d x } = \int_{\R^d} g(x) \pi(x)dx
\end{align*}
holds
with probability 1, for any initial state $\xi_0$ with $f(\xi_0)>0$.

How we choose the proposal distribution  $Q(\xi \to d \widetilde \xi)$ ? In fact we have almost complete freedom,
the only requirement is the mutual absolute continuity of the  1-step forward and backward measures:
\begin{align*}
\pi( \xi ) Q( \xi \to \xi') = 0  \Longleftrightarrow  \pi( \xi') Q( \xi' \to \xi )  =0 \; .
\end{align*}
In high dimension, to construct McMC proposals with good mixing properties can be very challenging and it is an art by itself.
A reference text is \citet{robert-casella}. 
A general idea is to update a subset of coordinates (block), keeping the rest fixed (Gibbs-Metropolis update). A Gibbs' update is a 
special case, where a subset of coordinates is updated
by sampling  a block from its conditional distribution  given  the remaining coordinates.
A Gibbs' update is always accepted.

In Bayesian inference, all the unknown parameters  and variables of the problem
  are thought as random variables with a given
prior probability distribution. Then the target distribution of the Metropolis-Hastings algorithm is the
posterior distribution of the unobserved variables conditionally on the observed ones. Bayes formula gives 
 \begin{equation*}
\mbox{Posterior$($ unobserved $|$  observed  $)  \propto $  Prior$($ unobserved $)\times$ 
Likelihood$($ observed $|$ unobserved  $)$}  
\end{equation*}
where  only the right hand side need to be specified and the normalizing constant  may remain unknown.

\subsection{Positivity constraints and McMC}
\label{sub:posi}

The  2nd-order tensor model in Eq. (\ref{eq:signal1}) describes the  decay of the 
signal $S_v(\bf q)$  in each direction ${\bf u }={\bf q}/{|{\bf q}| }$ as $| {\bf q}|$ increases.
In order to have physical meaning, the  diffusivity  function
$d({\bf u} ) = {\bf u}^T D_v {\bf u}$ should be non-negative,
 hence the  matrix $D_v$ must have non-negative eigenvalues.

In general, there are two simple ways to include a constraint $C \subset \R^d$ in a McMC algorithm.
In order to approximate the constrained expectation  
\begin{align*}
 E_{\pi}( g(\xi ) | \xi \in C) =\frac{  E_{\pi}( g(\xi) {\bf 1}( \xi \in C)  ) }
{ \pi(C ) } =\frac{  \int_{\R^d}  g(x) {\bf 1}_C( x) f(x) dx  }{  \int_{\R^d} {\bf 1}_C(x) f(x) dx },  
\end{align*}
One has to choose:
\begin{itemize}
 \item 
 include the constraint into the target distribution obtaining
a new target density proportional to $\widetilde f(x) = f(x) {\bf 1}_C(x)$. In practice this means
starting from a state $\xi_0\in C$, and  rejecting every proposed state which does not
satisfy the constraint. The resulting Markov chain takes values in the constraint set $C$.
\item alternatively,  include the constraint in the test function and sample from the   
unconstrained Metropolis-algorithm.
 By the law of large numbers, with probability 1
\begin{align*}
   E_{\pi}( g(\xi ) | \xi \in C) = \lim_{T\to\infty} \frac{
 \sum\limits_{t=0}^T  g( \xi_t) {\bf 1}_C(\xi_t)  }{ \sum\limits_{t=0}^T {\bf 1}_C( \xi_t) }\quad .
\end{align*} This second method has the advantage of simplicity, it is not even required to start the Markov chain from $\xi_0\in C$,
and the unconstrained Markov chain may have better mixing properties than the constrained one. The drawback  is 
 that the samples not satisfying the  constraint are lost.
\end{itemize}

\subsection{Bayesian hierarchical model} \label{sub:hierarchical}
We assign
 non-informative priors to the parameters of the likelihood function in Corollary 
 \ref{representation:coro}:
\begin{itemize}
\item
$\theta  \in \R^{d+1} $ has a flat shift-invariant improper prior $\pi(\theta)\propto 1$,
\item $\sigma^2$ has 
scale invariant improper prior, with density  $\pi(\sigma^2) \propto  1 / \sigma^2 $.

\end{itemize}
Given  the parameters $(\theta,\sigma^2)$, the random pairs $\{ (N_i,X_i): i=1,\dots,m \}$
are conditionally independent with conditional distribution
\begin{itemize}
\item $\bigl[ N_i  \big\vert \theta  ,  \sigma^2 \bigr]  \sim \mbox{Poisson}\biggl( 
 \exp( 2 \theta \cdot Z_i ) /( 2\sigma^2 ) \biggr)$,
\item $\bigl[ X_i \big\vert N_i, \sigma^2  \bigr] \sim$
 Gamma$\bigl( N_i+1 ,  1/   \bigl( 2 \sigma^2 \bigr) \bigr)$, $\quad Y_i= \sqrt{X_i}$. 
\end{itemize}

\subsection{Gibbs-Metropolis updates } \label{mcmc:update}
We combine sequentially several block updates, where in turn a subset of paramaters
is updated keeping the remaining ones fixed. When it is feasible, we sample the parameters from
their full conditional distribution (Gibbs' update). For the regression parameter 
$\theta$, we construct a Gaussian proposal distribution  which approximates the full conditional.
\begin{itemize}
\item  \textbf{Updating $\sigma^2$: } The variance parameter is updated in a Gibbs step. 
Conditionally on the augmented data $( N_i,Y_i,Z_i )$ and the parameter $\theta$,
the   conditional density of $\sigma^2$   up to a multiplicative constant is given by
\begin{align*}
 p( \sigma^2 | \theta,  N_i , Y_i, Z_i , i=1,\dots,m) \propto
 \exp\biggl( -  \frac{ 1 }{ 2\sigma^2} \sum_{i=1}^m  \bigl( Y_i^2 + \exp( 2 \theta\cdot Z_i)  \bigr)   \biggr)  (\sigma^2) ^{- \bigl
(1 + \sum\limits_{i=1}^m  (2N_i+1 ) \bigr) } 
\end{align*}
which corresponds to the inverse gamma distribution, with shape and rate parameters 
\begin{align*}
\sum_{i=1}^m  (2N_i+1 )  \quad \mbox{ and } \quad \frac 1 2
 \sum_{i=1}^m  \bigl( Y_i^2 + \exp( 2 \theta\cdot Z_i) \bigr) \;, \mbox{ respectively. } 
\end{align*}
\begin{rmk} Note that the noise variance $\sigma^2$ appears in both  augmented likelihood factors 
\begin{align*}   p( N_i | Z, \theta , \sigma^2 ) p( Y_i | N_i ,\sigma^2 )
\end{align*}
which makes the pair $( \theta_0,\sigma^2)$  identifiable. 
\end{rmk}

\item  \textbf{Updating $N$: } \label{N:update}
The auxiliary random variables $N_i$ are updated by sampling from the
full conditional distribution.
Conditionally on $\theta, \sigma^2 $ and the measurements $(Y_i, Z_i)$,  the  r.v's $N_i$
are conditionally independent  with reinforced Poisson distributions, with parameters
 \begin{align*}
\tau_i= Y_i\exp(Z_i \theta) /( 2\sigma^2) \;, ~ i = 1,\ldots, m, 
\end{align*}
respectively. In appendix A we discuss Monte Carlo sampling from the reinforced Poisson distribution.

\begin{rmk}  The augmented data $N$ is generated ``on the fly''
from the full conditional distribution above when needed. 
 It is not necessary to store $N$ into the computer memory.
\end{rmk}

 \item \textbf{Updating  $\theta$:} \label{theta:update}  
Conditionally on $N=(N_i: i=1,\dots,m)$ and $\sigma^2$, the parameter $\theta$
is independent of the observations $Y_i$, the
 full conditional distribution being proportional to 
\begin{align} \label{theta:logconditional}
 && p( \theta | \sigma^2, N )\propto \pi(\theta)\exp\biggl(\biggl( 2 \sum_{i=1}^m N_i Z_i \biggr) \theta
- \frac 1 {2\sigma^2 } \sum_{i=1}^m \exp( 2 Z_i \theta)  \biggr).
\end{align} 
Having assumed  a flat prior  $\pi(\theta)=1$,
we choose a Gibbs-Metropolis update with Gaussian  proposal distribution
\begin{align} \label{Gaussian:proposal}
          q(\theta|\hat\theta) \propto \exp\biggl(
 - \frac 1 2  ( \theta-\hat\theta)^{\top}  I ( \hat\theta) (\theta-\hat\theta)\biggr) \; ,
\end{align}
where have employed the Laplace approximation of Eq. (\ref{theta:logconditional})
 around the mode $\hat\theta$.
 Here  $\sigma^2$ and $N$ are fixed and the precision matrix is the Fisher information
\begin{align*} 
 I(\theta) = E_{\theta}\biggl( \nabla_{\theta} \log p( N | \theta , \sigma^2 )^{\top} \nabla_{\theta}
 \log p( N |\theta, \sigma^2 ) \biggr)
= \frac 2 { \sigma^2}  \sum_{i=1}^m \exp( 2 Z_i \theta) Z_i^{\top} Z_i.
\end{align*}
To find  the mode $\hat\theta$,
we use  the  iterative Fisher scoring algorithm \citep[see][]{mc_cullagh-nelder}, \citep{lange}Chapter 10.
The Hastings' ratio (HR) for $\widetilde \theta$ sampled from the proposal distribution $q(\cdot | \hat\theta)$ 
is given by
\begin{align*} 
   & \frac{ p( \widetilde \theta | \sigma^2,N )  q(\theta |\hat\theta)  }  { p( \theta | \sigma^2,N )  q(\widetilde \theta |\hat\theta )  } 
 =  & \\ &
  \exp\biggl(
 \biggl(\hat\theta^{\top} I(\hat\theta) - 2    \sum_{i=1}^m N_i Z_i  \biggr) (\theta-\widetilde \theta) +
  \frac 1 {2\sigma^2 } 
\sum_{i=1}^m\bigl\{  \exp( 2 Z_i \theta)- 
\exp( 2 Z_i \widetilde\theta) \bigr\}  + 
\frac 1 2 \widetilde\theta^{\top}  I(\hat\theta) \widetilde
\theta  - \frac 1 2 \theta^{\top} I(\hat\theta) \theta \biggr) \; . &
\end{align*}
\begin{rmk} Computing the Laplace approximation (Eq. \ref{Gaussian:proposal}) to the full conditional density
 (Eq. \ref{theta:logconditional}), is  crucial in order to get high acceptance rates in the McMC.
Without data augmentation, the GLM-likelihood in Eq. (\ref{theta:logconditional}) should be 
replaced by a product of Rice likelihoods. It is also possible 
to compute by Fisher scoring the  Laplace approximation of the full conditional
 under such Rice likelihood. However, for large sample size $m$,
 it could be  not computationally affordable to do that at every McMC update of every single tensor.
\end{rmk}

The algorithm is based on the assumption that the Fisher scoring algorithm converges to same global maximum $\hat \theta$
for all initial values $\theta$.
However, with a finite number of iterations, the approximate mode $\check \theta$ obtained by starting the 
Fisher scoring algorithm from the proposal  value $\widetilde\theta$  will be slightly different
than the approximate mode $\hat \theta$ obtained starting with  initial value $\theta$.  In order  to correct for this
discrepancy we have to run the Fisher scoring algorithm a second time starting from the proposed value $\widetilde \theta$
and reaching another approximate maximum $\check \theta$. In this case we redefine the Hastings' ratio as
\begin{align*}  \frac{ p( \widetilde \theta | \sigma^2,N )  q(\theta |\check\theta)  }  { p( \theta | \sigma^2,N )  q(\widetilde \theta |\hat\theta )  } 
 =&  \sqrt{ \frac{ \det  I(\check \theta)   }{ \det I(\hat \theta) } }  
    \exp\biggl(
 2\biggl(   \sum_{i=1}^m N_i Z_i  \biggr) (\widetilde \theta-\theta) +
  \frac 1 {2\sigma^2 } 
\sum_{i=1}^m\bigl\{  \exp( 2 Z_i \theta)- 
\exp( 2 Z_i \widetilde\theta) \bigr\}  \biggr)  & \\       & \times  \exp\biggl(
\frac 1 2  (\widetilde\theta -\hat \theta) ^{\top} I(\hat\theta) ( \widetilde \theta- \hat
\theta)- \frac 1 2 ( \theta - \check\theta) ^{\top}I(\check\theta) ( \theta- \check \theta)\biggr) &
\; .
\end{align*}

\begin{rmk} \label{rmk:fixingS_0}
Denote $S_0=S_v({\bf 0})$.
By fixing   $\theta_{0}=\log(S_0)$  to the current value, we can also update the tensor parameters $\theta_D\;$
 conditionally on $(\theta_{0}, \sigma^2,N)$.
This is useful in situations where data almost determine $S_0$ and the Fisher information  $I(\hat\theta)$
for $\theta=(  \theta_0 ,\theta_D)$
 is  numerically close to be singular.
In such cases Fisher scoring algorithm is unstable and may fail to converge. We take $\theta_0$ as known, 
and use instead  the Fisher information for $\theta_D$.
\end{rmk}
\item  \textbf{Separate update for $\theta_0$:} 
We consider also updating 
$\theta_{0}$ and the tensor $\theta_D$ separately.
We see that 
\begin{align*}
 p( N | \theta, \sigma^2 ) \propto ( S_0^2 )^a  \exp\bigl( - b S_0^2 \bigr)
\end{align*}
where 
\begin{align*}
  a=  \sum_{i=1}^m N_i , \quad b= \frac 1 {2\sigma^2}  \sum_{i=1}^m \exp\biggl( 2  Z_i  \left(
 \begin{matrix} 0 \\ \theta_D \end{matrix} \right) \biggr).
\end{align*}
Since $\log S_0$  has improper flat prior,   $\pi(S_0^2)\propto S_0^{-2}$  is the improper prior of $S_0^2$.
It follows that conditional on $( \theta_1,\dots,\theta_d), N$ and $ \sigma^2$,    $S_0^2$ 
is Gamma$(a,b)$-distributed.   We sample $\xi$ from this Gamma distribution and set $\theta_{0}= \log(\xi)/2$.
\end{itemize}

\subsection{Bayesian regularization of the tensor field } \label{dependence_prior}

Bayesian regularization is an image-denoising technique, introduced by  \cite{geman},
which has been already applied in DTI studies \citep{FHLVV,krissian}.
It is  assumed that under the prior distribution that the spatial parameters of the model
are not independent but form  a correlated random field. This is a reasonable assumption
in our context: even when a priori we do not have any information about the 
 main tensor direction at a given voxel, we know that often tensors from neighbour voxels are similar, just because a nervous fiber possibly continues from one voxel 
to the next. The prior dependence is taken into account according to  Bayes formula and it
has a smoothing and denoising effect on the posterior estimates.
An alternative,  is to estimate first  the parameters independently at each voxel, and then
 interpolate the preliminary tensor estimators to obtain a smoothed estimator.
The advantage of Bayesian regularization is  that estimation and regularization are performed
in a single procedure, by using all the available information.

Consider a zero mean $3\times 3$ symmetric Gaussian random matrix $D=( D_{i,j} : 1\le i \le j \le 3)$.
In \cite{basser-pajevic03},\cite{jeffreys}, it is shown
 that the distribution of $D$ is isotropic if and only if  it has  density of the form
\begin{align} \label{isotropic_prior:2}  &&
  p( D)= \frac{   \eta^{5/2} \sqrt{ \eta +3\lambda } }{ (\pi\sqrt 2)^3}
 \exp\biggl( - \frac 1 2  \biggl( \eta\mbox{Trace}( D^2 )+ \lambda  \{\mbox{Trace}(D) \bigr\}^2   \biggr) \biggr) 
\end{align}
with $\eta > 0$ and  $ \lambda > -\eta /3  $.  In  Section  \ref{SH:subsection}, we will see that (\ref{isotropic_prior:2})
follows from   the  isotropic Gaussian random field characterization  in terms of the law of its spherical harmonic coefficients.

For the vector $(D_{11} ,D_{22}, D_{33}, D_{12}, D_{13} ,D_{23 })$,
this corresponds to a Gaussian distribution with zero mean  and  precision matrix 
\begin{align}  \label{Omega_D:matrix:2nd}
\Omega_D= \left( \begin{matrix}\lambda + \eta & \lambda & \lambda & 0 & 0 & 0 \\
                     \lambda & \lambda + \eta & \lambda & 0 & 0 & 0 \\
\lambda & \lambda & \lambda + \eta  & 0 & 0 & 0 \\
0 & 0 & 0  & 2\eta & 0 & 0 \\
0 & 0 & 0  &0 & 2\eta & 0 \\
0 & 0 & 0 &0 &0 & 2\eta  \\
                    \end{matrix}
 \right).
\end{align}
We construct an (improper) pairwise-difference
Gaussian prior for a Markov random field of $(3\times 3 )$ symmetric matrices $(D(v) : v\in V)$ where $V$ is the set of voxels,
provided with the neighbourhood relation  $v\sim w$ in the $\Z^3$ lattice.
This Bayesian approach is equivalent to least-squares Tikhonov regularization in the framework of penalized maximum likelihood \citep{kaipio_somersalo}.
Define the improper prior density
\begin{align}\label{prior:2} 
 \begin{split} 
\pi( D(v) : v\in V)
\propto  &\exp\biggl( - \frac 1 {2 } \sum_{v\sim w }  \biggl(  \eta\mbox{Trace}(  \{ D(v)-D(w) \}^2 ) +
 \lambda \bigl\{  \mbox{Trace}( D(v)-D(w) ) \bigr\}^2 \biggr)
 \biggr)  =   \\ &
 \exp\biggl( - \sum_{v\sim w} \sum_{i=1}^3 \biggl\{ \frac{ (\eta+\lambda)} 2 ( D_{ii}(v)-D_{ii}(v) )^2 +  \\  &+ 
 \sum_{j< i} \biggl(  \lambda  ( D_{ii}(v)-D_{ii}(w) )( D_{jj}(v)-D_{jj}(v) ) + \eta ( D_{ij}(v)-D_{ij}(w))^2 \biggr) 
 \biggr\} \biggr) 
\end{split}
\end{align}
which is shift-invariant in $\R^6$ and invariant under rotations in $\R^3$.
The increments $(D(v)-D(w))$  have a proper rotation invariant
distribution, but the marginal prior of $D(v)$ does not integrate to a probability distribution.
For each voxel $v\in V$ introduce the regression parameter vector
\begin{align*}
\theta(v) & = \bigl(\theta_0(v),\theta_1(v),\theta_2(v),\theta_3(v),
  \theta_4(v), \theta_5(v),\theta_6(v)\bigr) &
\\ &= \bigl
(       \log( S_0(v)) ,   D_{11}(v),D_{22}(v),D_{33}(v),D_{12}(v),D_{13}(v),D_{23}(v)  \bigr). &
\end{align*}
For the  log-intensity parameters $\theta_0(v)= \log( S_0(v) )$ we could either assume prior independence and assign
 a flat  prior,  or use a pairwise
difference improper contextuality prior with density
\begin{align*}
  \pi( \theta_0 (v): v\in V) \propto\exp\biggl( - \frac{ \rho }{ 2} \sum_{v\sim w} \bigl( \theta_0(v)-\theta_0(w) \bigl)^2 \biggr) \; ,
\end{align*}
called instrinsic prior \citep{besag_york_moille}.
The hyperparameters $\eta,\rho\ge 0$,  $\lambda> -\eta/3$, are tuning the correlations of the difference $(\theta(v)-\theta(w))$. 
As in Section \ref{data:augmentation:section},  for each
 voxel $v$   we introduce:
\begin{itemize}
\item
a  noise-parameter $\sigma^2(v)>0$ with scale-invariant improper prior $\propto \bigl( \sigma^2 (v) \bigr)^{-1}$,
\item
a random vector $N(v)=(N_k(v): k=1,\dots,m)$ which follows the generalized linear model of Corollary \ref{representation:coro}
with Poisson response distribution and logarithmic link function, covariate matrix $Z\in m \times (d+1)$ and parameter  $\theta(v)$.
\end{itemize}
Here $(\sigma(v):v\in V)$ are independent and $(N(v):v\in V)$ are conditionally independent given $(\theta(v):v\in V)$.

As before, we compute the  Laplace approximation for the log-likelihood at each voxel $v$.
 When we combine this Gaussian log-likelihood approximation  with the 
pairwise-difference Gaussian prior by using Bayes formula, we obtain
 an approximating Gaussian posterior for $\theta(v)$, which we will use as 
proposal distribution in the Gibbs-Metropolis update.
\nolinebreak
\nolinebreak
We may consider 
the single site update, where $\theta(v)$ is updated voxelwise conditionally
on  $N(v)$ and the values $\theta(w)$ at neighbour voxels $v\sim w$.
Alternatively we can construct a Gaussian approximation to the full conditional as a joint proposal
in a  simultaneous update for a block $( \theta(v)\in W )$, where $W\subseteq V$ is a connected subset of voxels.  
The size of a block can vary from a single site to the whole brain. 
For example  we may define a block as  a ball with given center and radius
under  the graph distance, which 
is the length of the shortest path between two voxels.
We denote the exterior boundary  of $W$ by
\begin{align*}
 \partial W := \{  w \in V\setminus  W : \; \exists v\in W  \mbox{ with } w\sim v\}
\end{align*}
and set $\overline W:= W \cup \partial W $,
 $\partial\{ v\}:=\{ w \in V : w\sim v \}$ denotes the neighbourhood of $v$, and $\#\partial\{ v\}$ stands for its cardinality.
We update the variable $(\theta(w) : w \in W )$ conditional on the observations  $(N(w): w\in  W)$ and $(\theta(v): v\in \partial W)$.

The prior of $(\theta(w): w \in W \cup \partial W )$ is Gaussian and the likelihood of $\theta(w)$ with respect to the
augmented data  $N(w)$ 
is approximated by the Gaussian density ${\mathcal N}( \hat\theta(w), \hat I(w)^{-1} )$,
where $\hat\theta(w)$ and $\hat I(w)$ are functions of $N(w),\sigma^2(w)$  and the design matrix $Z$,
 computed by using Fisher scoring under the Poisson GLM
 as in Section  \ref{theta:update}.
The corresponding  Gaussian posterior   distribution  $q( \theta(w) : w \in W ) $ will be used as proposal in the  Metropolis block update,
and satisfies
\begin{align*} 
\log q( \theta(w) : w \in W ) =&
 \mbox{ const.} - \frac 1 {2 } \sum_{  w\sim v : v\in W, w \in \overline W    }  \biggl(  \eta\mbox{Trace}(  \{ D(v)-D(w) \}^2 ) +
 \lambda \bigl\{  \mbox{Trace}( D(v)-D(w) ) \bigr\}^2  \biggr) &\\ &
- \frac {\rho}{ 2} \sum_{  w\sim v : v\in W, w \in \overline W    } (\theta_0(v)-\theta_0(w))^2
- \frac 1 2 \sum_{v\in W } ( \theta(v)-\hat\theta(v)) ^{\top}\hat I(v) ( \theta(v)-\hat\theta(v)) & \\
=& \; \mbox{const.} 
- \frac 1 2  \sum_{v\in W} \theta(v) ^{\top}\biggl(  \# \partial\{ v\}   \Omega   + \hat I(v) \biggr) \theta(v)
+ \sum_{v\sim w :  v,w\in W } 
\theta(v)^{\top}  \Omega \theta(w) & \\ &
+
\sum_{v\in W } 
\theta(v)^{\top} \biggl( \hat I(v) \hat \theta(v) + 
\Omega \biggl( \sum_{w\in \partial\{ v\}\setminus W  } \theta(w) \biggr) \biggr)
&\\ = & \; \mbox{const.}  - \frac 1 2 \sum_{v,w \in W : \;w=v\;\mbox{\small or }w\sim v } ( \theta(v)-\hat \mu(v) ) ^{\top} \; 
\hat \Psi_{v,w}\; (\theta(w)-\hat \mu(w) ) \;, &
\end{align*}
where the constant term 
does not depend on $(\theta(v): v\in W)$ and may change from line to  line,
\begin{align} \label{omega:matrix} 
\Omega= \left( \begin{matrix}  \rho & 0 
 \\  0 & \Omega_D
                    \end{matrix}
 \right)
\end{align}
is a  $7\times 7$ precision matrix,
and after completing the squares we have defined
\begin{align*} 
 \mu^{\top}  =&(\hat  \Psi)^{-1} \hat \xi^{\top} \quad\mbox{ with }\quad
 \hat
   \xi(v)^{\top}= \hat I(v) \hat \theta(v) + 
 \Omega \biggl( \sum_{w\in \partial\{ v\} \setminus W  } \theta(w)\biggr)\quad \mbox{ and } & \\
\hat \Psi_{v,w} = & \biggl(  \# \partial\{ v\}{\bf 1}( v=w ) -{\bf 1}( v\sim w)  \biggr)  \Omega
 + {\bf 1}( v = w ) \hat I(v) \; ,&
 \end{align*}
is a  band diagonal precision matrix with  $(7\times 7)$ blocks and $ v,w\in W$.
This corresponds to a Gaussian proposal distribution  $q(\theta(w):w\in W)$ 
with mean $( \hat   \mu(w) : w\in W)$ and covariance 
$(\hat \Psi)^{-1}$.
\paragraph{Prior contribution} 
The prior contribution is derived as the proposal contribution by  
conditioning on the values $(\theta(v): v\in \partial W )$ without including data. We obtain
\begin{align*} 
\log \pi( \theta(w) : w \in W ;  \theta(v), v\in \partial W)  & =
 \mbox{const.} 
-\frac 1 2 \sum_{ v\sim w :v \in W,w\in \overline W } (\theta(v)-\theta(w) ) ^{\top} \Omega  (\theta(v)-\theta(w) )& \\ &
=\mbox{const.} - \frac 1 2 \sum_{v,w\in W} \theta(v) ^{\top} \Phi_{v,w} \theta(w) +\sum_{v\in W } 
\theta(v)^{\top}   \Omega \biggl( \sum_{w\in \partial\{ v\}\setminus W  } \theta(w) \biggr)
&\\ \mbox{with}  \quad\quad
 \Phi_{v,w}: & = \biggl(  \# \partial\{ v\}{\bf 1}( v=w ) -{\bf 1}( v\sim w)  \biggr)  \Omega , \quad v,w\in W. & 
\end{align*}
These expressions   determine the Hastings' ratio for this Gibbs-Metropolis update (here omitted).

\subsection{Updating the regularization parameters of the 2nd order tensor field} \label{regularization}

The precision matrix of the  Gaussian random field $( \theta(v): v\in V)$ is the  Kronecker
product $\Gamma\otimes \Omega_D$, where $\Gamma_{v,w}= \Gamma_{v,w}= {\bf 1}( v\sim w)$
is the adjacency matrix of the graph $V$, and $\Omega_D$ was given in  (\ref{Omega_D:matrix:2nd}). 
Since 
\begin{align*}
   \det( \Gamma \otimes \Omega_D )=  \det( \Gamma)^6 \det( \Omega_D)^{|V|Â } \; .
\end{align*}
 the likelihood for $\lambda, \eta$ based on 
 $\bigl(\theta(v): v\in V\bigr)$ is proportional to
\begin{align*}  \propto
 \bigl(  \eta^{5/2} \sqrt{ \eta +3\lambda } \bigr )^{|V|}
  &\exp\biggl( - \frac 1 {2 } \sum_{v\sim w }  \biggl(  \eta\mbox{Trace}(  \{ D(v)-D(w) \}^2 ) +
 \lambda \bigl\{  \mbox{Trace}( D(v)-D(w) ) \bigr\}^2 \biggr)
 \biggr)  \;,
\end{align*}
with constraints $\eta > 0$ and  $\lambda > -\eta/3$.

In order to factorize the likelihood we reparametrize with $\delta=(\eta+3\lambda)$, obtaining
\begin{align*}
 &\eta^{|V |5/2} 
\exp\biggl( - \eta \sum_{v\sim w } \biggl(
\frac 1 2 
 \mbox{Trace}(  \{ D(v)-D(w) \}^2 )-\frac 1 6     \bigl\{ \mbox{Trace}( D(v)-D(w) ) \bigr\}^2       \biggr) 
\biggr)
& \\ & \times \; \delta ^{|V|/2 }\exp\biggl( - \frac {\delta} {6 }  \sum_{v\sim w }  \bigl\{ \mbox{Trace}( D(v)-D(w) ) \bigr\}^2 \biggr)
\; .&
\end{align*}
Assuming scale invariant independent priors for $\eta,\delta$,
\begin{align*}
  \pi( \delta,\eta ) \propto \delta^{-1} {\bf 1}( \delta  >0 ) \; \times \; \eta^{-1} {\bf 1}( \eta >0 )
\end{align*}
we obtain the full conditional distribution of  $(\delta,\eta)$ 
as the product of two Gamma densities, 
\begin{align*} &
 \pi( \delta |  \theta )\sim \mbox{ Gamma}\biggl( \frac{ |V| } 2,\frac 1 6  \sum_{v\sim w }  \bigl\{ \mbox{Trace}( D(v)-D(w) ) \bigr\}^2  \biggr) 
&  \\ &
\pi( \eta | \theta)\sim \mbox{ Gamma}\biggl(  \frac{|V| 5} 2 ,   \sum_{v\sim w } \biggl(
 \frac 1 2 \mbox{Trace}(  \{ D(v)-D(w) \}^2 )-\frac 1 6     \bigl\{ \mbox{Trace}( D(v)-D(w) ) \bigr\}^2
  \biggr) \biggr) \; .
\end{align*}
In the McMC, we update the regularization parameters by sampling
 $(\eta,\delta)$ independently from these full conditional distribution and setting $\lambda=(\delta-\eta) /3$.


\subsection{ Modeling diffusivity with 4th-order tensors} \label{sub:4thorder_model}
Several authors,
 (\cite{basser-pajevic07,Mori,ghosh,moakher,GPD-inria}), 
argue that the 2nd-order tensors
fail to capture complex tissue structures such as fibers crossing and branching
 in a single voxel. In such voxels most often  anisotropy is underestimated
and fiber tracking  algortihms based on 2nd-order tensors estimates terminate.
In fact, while at every spatial location we have a diffusion matrix, in the time
scales we are considering, the scale 
of water diffusion is of smaller order than the size of a voxel.
 The 2nd-order tensor model assumes 
that the diffusion tensor is constant at all points inside one voxel.  
In reality a voxel contains a whole population
of cellular structures, corresponding to a population of diffusion tensors.
Equation $(\ref{eq:signal1})$ should be replaced by
\begin{align}\label{mixed:gaussian}\frac{  S_v({\bf q})}{ S_v({\bf 0 }) } \; = \;   E_{v}\biggl( \exp\bigl( 
i \;{\bf q} \cdot {\bf V}_t 
 \bigr ) \biggr) 
\; = \; 
\int\limits_{{\mathcal M}^+ }
\exp\biggl( - \frac 1 2  
{\bf q}^{\top}  D  {\bf q }
\biggr) d Q_{v}( D )\; ,\end{align}
which is the characteristic function
of the random displacement  ${\bf V}_t$  of a water molecule
randomly selected within the voxel. 
Here  $Q_v$ is a probability distribution
on the space ${\mathcal M}^+\subset\R^{6\times 6} $ of positive definite matrices 
for the population of diffusion tensors.
Instead of measuring the characteristic function of centered Gaussian
random vector, the MR-experiment measures the characteristic function of a  Gaussian mixture.
We see from (\ref{mixed:gaussian})  that  the signal
 $S_v(  {\bf q}    )$ is a decreasing function of $|{\bf q}|$.
In $4$-th order tensor modeling it is assumed that 
the signals are given by 
\begin{align} \label{diffusivity:4}
  S_v( {\bf q} )=   S_v( {\bf 0} )\exp\bigl( - b d({\bf u})  \bigr)= \exp( Z\theta), \quad \quad {\bf q}\in \R^3,
\end{align}
where  $b=|{\bf q}|^2/2$ is the $b$-value, ${\bf u}= {\bf q}/|{\bf q}|$ is the  gradient direction,
and the {\it  diffusivity function } 
\begin{align} \label{diffusivity}
 d({\bf u})=
D:( {\bf u }\otimes {\bf u} \otimes {\bf u} \otimes {\bf u} )  : = 
  \sum_{i_1=1}^3 \sum_{i_2=1}^3 \sum_{i_3=1}^3 \sum_{i_4= 1}^3 D_{i_1 i_2 i_3 i_4}  u_{i_1} u_{i_2} u_{i_3} u_{i_4}, \quad  {\bf u}\in S^2
,
\end{align}
 is an homogenous polynomial of degree $4$.  Here the 4-th order tensor
\begin{align*}D=\bigl( D_{i_1 i_2 i_3 i_4}:  1\le i_1\le i_2 \le i_3 \le i_4 \le 4  \bigr)\end{align*}
is totally symmetric.  In (\ref{diffusivity:4}) we have introduced the parameter $\theta\in \R^{15}$ as
\begin{align*} &\bigl(\log S({\bf 0}),D_{1111}, D_{2222},D_{3333},D_{1122},D_{1133}, D_{2233}, 
D_{1123},D_{1223}, D_{1233},
D_{1112},D_{1113}, D_{1222}, D_{2223} ,D_{1333},D_{2333}\bigr)^{\top},
\end{align*}
and the design matrix $Z= \bigl({\bf 1}^{\top} ,Z_D\bigr ) \in \R^{m\times 15} $ with rows\begin{align*}
Z_D=-(  u_1^4 ,u_2^4 ,u_3^4,  6 u_1^2 u_2^2 , 6 u_1^2 u_3^2, 6 u_2^2 u_3^2,
12 u_1^2 u_2 u_3, 12  u_2^2 u_1 u_3,  12 u_3^2 u_1 u_2,
4 u_1^3 u_2, 4 u_1^3 u_3, 4 u_2^3 u_1,4 u_2^3 u_3, 4 u_3^3 u_1, u_3^3  u_2 ) b.
\end{align*}
Because the diffusivity function models signal decay,
the $4$-th order tensor must satisfy the positivity constraint
\begin{align*}
D:( {\bf u }\otimes {\bf u} \otimes {\bf u} \otimes {\bf u} ) \ge 0,  \quad \forall {\bf u}\in S^2.
\end{align*}
When we analyze the data at each voxel separately,
under  the  high order tensor diffusivity model, only the dimensions of the parameter $\theta$ and the design matrix $Z$ are changed, and
the data augmentation of Section \ref{data:augmentation:section} and the Bayesian procedures of Section
 \ref{bayes:modeling} apply directly.

In what follows,
in order to perform Bayesian regularization of the tensor field, we first give 
the  general form of an isotropic Gaussian distribution for the $4$-th order tensor, 
in analogy with  (\ref{isotropic_prior:2}).
 Then, by taking pairwise differences, we obtain an isotropic  Gaussian random field of $4$-th order tensors
which replaces the prior (\ref{prior:2}) in  the Bayesian regularization method
of  Section \ref{dependence_prior}.
\\

In \cite{basser-pajevic07}, 
the 4th-order tensor in dimension $3$  is shown to be isomorphic to a  2nd-order tensor in dimension $6$ under the isomorphism
\begin{align}\label{Dhat}
 D\longmapsto  \widehat D:= \left( \begin{matrix} D_{1111} & D_{1122}&  D_{1133} & \sqrt{2} D_{1112}  & \sqrt 2 D_{1113} &\sqrt 2 D_{1123} \\
       D_{1122} & D_{2222}&  D_{2233} & \sqrt{2} D_{1222}  & \sqrt 2 D_{1223} &\sqrt 2 D_{2223}  \\
           D_{1133} & D_{2233}&  D_{3333} & \sqrt{2} D_{1233}  & \sqrt 2 D_{1333} &\sqrt 2 D_{2333} \\
     \sqrt 2 D_{1112}  & \sqrt 2  D_{1222} & \sqrt{2} D_{1233} & 2 D_{1122} & 2 D_{1123} & 2 D_{1223} \\
  \sqrt 2 D_{1113}  & \sqrt 2  D_{1223} & \sqrt{2} D_{1333} & 2 D_{1123} & 2 D_{1133} & 2 D_{1233} \\
 \sqrt 2 D_{1123}  & \sqrt 2  D_{2223} & \sqrt{2} D_{2333} & 2 D_{1223} & 2 D_{1233} & 2 D_{2233}  
                     \end{matrix} 
\right).
\end{align} 
The six eigenvalues and eigentensors of the $4$-th order tensor $D$, correspond to the eigenvalues and eigenvectors of the matrix $\widehat D$.
Furthermore, it is shown in \cite{GPD-inria}, that
$\mbox{ Trace}( \widehat D)^2,\mbox{Trace}( \widehat D^2 ) $ and the polynomial
\begin{align} \label{g:polynomial}
\begin{split} 
 g(D) & = D_{1111} ( D_{2222}+D_{3333} ) + D_{2222}D_{3333} + 
3  \biggl\{ D_{1122}^2  + D_{1133}^2 + D_{2233}^2 \biggr\} \\ 
&+2 \biggl\{  D_{1122} D_{3333} + D_{1133}D_{2222} + D_{2233} D_{1111} +
D_{1122}( D_{1133} + D_{2233} ) + D_{2233}D_{1133} \biggr\}
\\ &+
4 \biggl\{   D_{1233} ( D_{1233} -D_{1222} - D_{1112}  )
+  D_{1223} (D_{1223} -D_{1113} - D_{1333}  ) \\ 
&+  D_{1123} ( D_{1123} -D_{2333} - D_{2223} ) 
- D_{1222}D_{1112} - D_{1113}D_{1333} - D_{2223}D_{2333} \biggr\} 
\end{split}
\end{align}
are invariant under 3d-rotations  and span the space of isotropic
homogeneous polynomials of degree 2 in the variables $D$. 
Here we give the general form of a zero-mean isotropic Gaussian distribution for the 4th-order tensor, with density
\begin{align} \label{rotation:invariant4th} 
  \pi( D) =  2^3 
\sqrt{   \frac{  (\gamma+\eta)^{9} ( 3\eta-4\gamma)^{5} ( 3\eta + 8\gamma + 15\lambda ) } {\pi^{15} } }
\exp\biggl( - \frac 1 2 \biggl\{  
\eta \mbox{Trace}( \widehat D^2 ) + \lambda \mbox{ Trace}( \widehat D)^2    + \gamma g(D)\biggr\}  \biggr) \; .
\end{align}
Again (\ref{rotation:invariant4th}) follows from the characterization of isotropic Gaussian
random fields in terms of the law of their spherical harmonic coefficients, which we discuss in Section 
 \ref{SH:subsection}.

Under  (\ref{rotation:invariant4th}), the random coefficients   
$(D_{1111}, D_{2222} , D_{3333}, D_{1122} ,D_{1133} ,D_{2233})$  have precision matrix  
\begin{align*}
\Omega^{'}=\left(\begin{matrix}  \eta+ \lambda  & \lambda + \gamma & \lambda + \gamma &  2 \lambda  & 2 \lambda  & 2 \lambda +  2\gamma  \\
 \lambda  + \gamma & \eta + \lambda & \lambda  +\gamma &  2 \lambda  & 2 \lambda + 2\gamma & 2 \lambda \\
 \lambda  + \gamma &  \lambda +\gamma & \eta +\lambda  &  2 \lambda +2 \gamma  & 2 \lambda  & 2 \lambda  \\
 2 \lambda &  2 \lambda & 2\lambda +2\gamma& 6 \eta+ 6\gamma+ 4 \lambda & 4 \lambda + 2\gamma & 4 \lambda + 2\gamma  \\     
2 \lambda &  2 \lambda + 2\gamma & 2\lambda &  4 \lambda +2 \gamma & 6\eta + 6\gamma + 4 \lambda & 4 \lambda+ 2\gamma\\
2 \lambda  + 2\gamma &  2 \lambda & 2\lambda & 4 \lambda +2 \gamma & 4 \lambda  + 2\gamma & 6 \eta+ 6\gamma + 4 \lambda
     \end{matrix}
 \right)\; ,
\end{align*}
and are independent from
$( D_{1112},D_{1113}, D_{1222}, D_{2223} ,D_{1333},D_{2333} D_{1123},D_{1223}, D_{1233} )$,
which have precision matrix
\begin{align*} && \Omega^{''}=
\left(\begin{array} {ccccccccc}
              \\       
     4 \eta   &      0   & -4 \gamma  & 0 & 0& 0 &0 &0 &-4 \gamma     \\
     0   &      4\eta   & 0   & 0 & -4  \gamma   & 0 &0 &-4 \gamma    &0       \\
     -4   \gamma    &      0   & 4\eta   & 0 & 0& 0 &0 &0 &-4  \gamma        \\
     0   &      0   & 0   & 4 \eta & 0& -4  \gamma   &-4  \gamma   &0 &0       \\
     0   &      -4  \gamma     & 0   & 0 & 4 \eta & 0 &0 &-4  \gamma   &0       \\
     0   &      0   & 0   & -4  \gamma   &0& 4\eta &-4  \gamma   &0 &0       \\
     0   &      0   & 0  & -4  \gamma   & 0 & -4  \gamma   & 12\eta +8  \gamma   &0 &0       \\
     0   &      -4 \gamma & 0   & 0 & -4 \gamma    & 0 &0 & 12\eta +8 \gamma    &0       \\
     -4  \gamma    &      0   & -4 \gamma      & 0 & 0& 0 &0 &0 &12 \eta +8  \gamma       
\end{array} 
\right)   \;.
\end{align*}
The covariance matrix of $D$ is positive definite under the constraints
\begin{align*}
 \eta>0,  \quad \frac 3 4 \eta > \gamma > -\eta, \quad \lambda > - \biggl(\frac 1 5  \eta + \frac 8 {15}  \gamma \; \biggr)\; .
\end{align*}
The construction and block-updates described in Section \ref{dependence_prior}  extends
 directly to a $4$-th order tensor valued random field $( D(v):v\in V)$, 
with the improper rotation-invariant pairwise-difference Gaussian prior
\begin{align*} &
\pi( D(v) : v\in V) 
\propto  & \\ &
\exp\biggl( - \frac 1 {2 } \sum_{v\sim w }  \biggl(  \eta\mbox{Trace}(  \{ \widehat D(v)- \widehat D(w) \}^2 ) +
 \lambda \bigl\{  \mbox{Trace}( \widehat D(v)-\widehat D(w) ) \bigr\}^2
+ \gamma  g( D(v)-D(w) ) \biggr)
 \biggr)   \; .& 
\end{align*}
Inside the exponential, appears a generalization of  the  regularization  term  used in \citet{barmpoutis2}.
In order to proceed as in  Section \ref{dependence_prior},
we just need to replace the precision matrix in   (\ref{omega:matrix}) by the $16\times 16$ block-diagonal matrix
\begin{align} \label{Omega_D:matrix:4th}
\Omega= \left(  
\begin{matrix} \rho  &  0  & 0 \\  
   0 &  \Omega^{'} & 0  \\
  0  & 0 & \Omega^{''}
\end{matrix}
\right ) \; .
\end{align}

\paragraph{Positivity constraint for  4th order tensors.} It follows that the diffusivity function $d({\bf u})$ in (\ref{diffusivity}) is positive
when the $6\times 6$ matrix $\widehat D$ in (\ref{Dhat})  has positive eigenvalues.
This is a sufficient but not a necessary condition, because it is enough to
have positivity on the algebraic variety
\begin{align*}
  \bigl \{  ( u_1^2, u_2^2, u_3^2, u_1 u_2, u_1 u_3 , u_2 u_3 ) :  \;  (u_1, u_2, u_3) \in \R^3 \bigr\} \subset \R^6  \; .
\end{align*}
When $\widehat D$ is negative definite, we should check the sign of the $Z$-eigenvalue of the diffusivity, 
which was introduced by \cite{qi}
as the solution of the constrained optimization problem
\begin{align*}
\lambda=\min \bigl\{  d( {\bf u}) : \;   {\bf u}  \in \R^3, \;  |{\bf u} | = 1 \bigr\} \; .
\end{align*}


\subsection{ Updating the parameters of the 4th-order tensor field }\label{sub:update:regularization_4}

 The likelihood for $\lambda, \eta,\gamma$ based on 
 $\bigl(\theta(v): v\in V\bigr)$ is proportional to
\begin{align*} &  \propto {\bf 1}\bigl( \eta > 0 \bigr) {\bf 1}\bigl(  3/4 \eta > \gamma > -\eta \bigr) {\bf 1}\bigl( \lambda + \eta/5 + \gamma 8/15  > 0\bigr) 
 \biggl\{Â    (\gamma+\eta)^{9} (3\eta-4\gamma)^5 ( 3\eta +8 \gamma + 15\lambda)  \biggr \}^{|V|/2}
 & \\ &\exp\biggl( - \frac 1 {2 } \sum_{v\sim w }  \biggl(  \eta\mbox{Trace}(  \{ \widehat D(v)-\widehat D(w) \}^2 ) +
 \lambda \bigl\{  \mbox{Trace}( \widehat D(v)-\widehat D(w) ) \bigr\}^2 +  \gamma g\bigl( D(v)-D(w) \bigr) \biggr)
 \biggr)  &
\end{align*}
where  the polynomial  $g(D)$ was given in (\ref{g:polynomial}). In order to factorize the likelihood we reparametrize with 
\begin{align*} 
  \alpha=( \gamma+\eta), \; \beta= (3\eta-4\gamma) , \;\delta=( 3\eta +8 \gamma + 15\lambda) 
\end{align*}
with $\alpha, \beta, \delta > 0$. The linear system  has solution
\begin{align} \label{linear solution:4th order}
\eta= \frac{ \beta + 4 \alpha } 7, \; \lambda=\frac{  7 \delta+ 5 \beta - 36 \alpha } { 105},\;\gamma=\frac{  3 \alpha - \beta }  7 , 
\end{align}
and the corresponding likelihood is proportional to
\begin{align*} & 
\alpha^{ | V| 9/2 } \exp\biggl(  - \frac {\alpha}  {14} 
\sum_{v \sim w} \biggl\{Â   4  \mbox{Trace}(  \{ \widehat D(v)-\widehat D(w) \}^2 - \frac{ 12 }{ 5} 
\bigl\{  \mbox{Trace}( \widehat D(v)-\widehat D(w) ) \bigr\}^2 +3 g\bigl( D(v)-D(w) \bigr) \biggr\} \biggr) \times \\ & 
\beta^{ |V| 5/2 } 
\exp\biggl(  - \frac {\beta}  {14} 
\sum_{v \sim w} \biggl\{Â    \mbox{Trace}(  \{ \widehat D(v)-\widehat D(w) \}^2  )
+ \frac 1 {3 }\bigl\{  \mbox{Trace}( \widehat D(v)-\widehat D(w) ) \bigr\}^2 -g \bigl( D(v)-D(w) \bigr) \biggr\} \biggr)  \times
& \\ &
\delta^{ |V|/2}
\exp\biggl(  - \frac {\delta}  {30}
\sum_{v \sim w}
 \bigl\{  \mbox{Trace}( \widehat D(v)-\widehat D(w) ) \bigr\}^2\biggr)  \; .
&
    \end{align*}
We assume scale invariant 
priors for $\alpha,\beta,\delta$,
\begin{align*}
  \pi( \alpha,\beta,\delta) \propto \alpha^{-1} {\bf 1}( \alpha >0 ) \; \times \;  \beta^{-1} {\bf 1}( \beta >0 ) \; \times \; 
 \delta^{-1} {\bf 1}( \delta  >0 ) \; , 
\end{align*}
and obtain the full conditional distribution of  $(\alpha,\beta,\delta)$ 
as the   product of these Gamma densities: 
\begin{align*} &
\pi( \alpha | \theta) \sim \mbox{Gamma}\biggl( \frac 9 2  |V| , \; \frac 1  {14} 
\sum_{v \sim w} \biggl\{Â   4  \mbox{Trace}(  \{ \widehat D(v)-\widehat D(w) \}^2 - \frac{ 12 }{ 5} 
\bigl\{  \mbox{Trace}( \widehat D(v)-\widehat D(w) ) \bigr\}^2 +3 g\bigl( D(v)-D(w) \bigr) \biggr\} \biggr) \biggr)  &
\\&
\pi( \beta | \theta) \sim \mbox{Gamma}\biggl(\frac 5 2 |V| , \;  \frac 1 {14}
\sum_{v \sim w} \biggl\{Â    \mbox{Trace}(  \{ \widehat D(v)-\widehat D(w) \}^2  )
+ \frac 1 {3 }\bigl\{  \mbox{Trace}( \widehat D(v)-\widehat D(w) ) \bigr\}^2 -g \bigl( D(v)-D(w) \bigr) \biggr\} \biggr)  \biggr)
& \\&
\pi( \delta | \theta) \sim \mbox{Gamma}\biggl(  \frac{Â |V|}   2, \;  \frac 1 {30} \sum_{v \sim w}
 \bigl\{  \mbox{Trace}( \widehat D(v)-\widehat D(w) ) \bigr\}^2 \biggr)  \; .&
\end{align*}
In the McMC, $(\alpha,\beta,\delta)$ are updated independently by sampling from these full conditionals.
The corresponding $(\eta, \lambda, \gamma)$  are then obtained from equation (\ref{linear solution:4th order}).

\subsection{  Spherical harmonics representation } \label{SH:subsection} In general,
the  diffusivity function  $d:S^{2}\to \R$ can be expanded as 
\begin{align}  \label{diffu:expansion}
   d(u)=\sum_{\ell \in 2\N } \sum_{m=-\ell}^{\ell} \theta_{\ell,m} Y_{\ell,m}(u) , \quad  u\in S^2
\end{align}
where
\begin{align*}
  \theta_{\ell,m} = \bigl\langle  d, Y_{\ell,m} \rangle_{L^2(S^2) }:=\int_{ S^2}  d(u) Y_{\ell,m} (u) \sigma(du) \; ,
\end{align*}
and the  {\it real spherical harmonics} $(Y_{\ell,m}(u): \; \ell \in \N, m=-\ell, \dots, \ell)$
are homogeneous polynomials of respective degrees $\ell$ 
 forming an orthonormal basis of  $L^2( S^2, d\sigma)$  equipped with the Haar measure $\sigma(du)$
(see \cite{peccati:marinucci}, Paragraph 3.4).
Because of  the symmetry $d(u)=d(-u)$ $\forall u \in S^2$, only the spherical harmonics of even degree contribute to  (\ref{diffu:expansion}).
By  truncating the expansion  (\ref{diffu:expansion}) up to polynomials of degree $2n$, 
we obtain a finite dimensional parametrization,
which corresponds to the tensor model of order $2n$
\begin{align} \label{tensor:representation}
 d(u)= \sum_{i_1=1}^3 \dots  \sum_{i_{2n}=1}^3 D_{i_1 \dots i_{2n}  } u_{i_1}  \dots u_{i_{2n}}
=\sum\limits_{\kappa\in \N^3: |\kappa|=2n} \mu_{\kappa}D_{\kappa}  u_1^{\kappa_1}u_2^{\kappa_2}u_3^{\kappa_2} \; ,
\end{align}
 where the tensor $D$ is totally symmetric and the coefficients $D_{\kappa}$ have multiplicities
\begin{align*}
\mu_{\kappa}=\frac{ |\kappa|!}{ \kappa_1! \kappa_2 ! \kappa_3! },  \quad |\kappa|=\sum\limits_{i=1}^3 \kappa_i = 2n\; .
\end{align*}
 By comparing
the representations  (\ref{tensor:representation}) and
(\ref{diffu:expansion}) as in \cite{ozars}, it follows that the coefficients of
the tensor of order $2n$ and the spherical harmonic coefficients
of degrees $0,2,\dots,2n$ are related by a linear bijection  $D=\theta B$.
For the 2nd-order tensor model this holds for
\begin{align*} & D= \bigl( D_{11} ,D_{22} , D_{33}, D_{12} ,D_{13}, D_{23} \bigr ), & \\ &
\theta=\bigl( \theta_{0,0}, \theta_{2,-2}, \theta_{2,-1}, \theta_{2,0},\theta_{2,1},\theta_{2,2} \bigr) , \; 
B= \left(\begin{matrix}
  \frac 2{\sqrt{15}}  &  \frac  2{\sqrt{15}} &  \frac{2 }{\sqrt{15}} & 0 & 0 & 0 \\
   0  &  0 &  0 &  1 &  0 &  0 \\
              0  &  0 &  0 &  0 &  0 &  1 \\ 
  - \frac{ 1 }{ \sqrt 3}  & -\frac 1  {\sqrt 3}  &  \frac{  2 }{\sqrt 3}  &   0  &  0   &   0 \\
   0  &  0 &  0 &  0 &  1 &  0 \\ 
   1   &  -  1 & 0  &  0   &   0   &   0 
         \end{matrix}
\right)\frac 1 4 \sqrt{\frac{15}{ \pi} } \;,&
\end{align*}
and for the 4th order tensor model it holds with
\begin{align*}
 &
\theta=\bigl( \theta_{0,0}, \theta_{2,-2}, \theta_{2,-1}, \theta_{2,0},\theta_{2,1},\theta_{2,2},
\theta_{4,-4}, \theta_{4,-3},\theta_{4,-2}, \theta_{4,-1} , \theta_{4, 0},
\theta_{4,1}, \theta_{4,2},\theta_{4,3}, \theta_{4,4} \bigr) ,
\\  &
 D= \bigl( D_{1111} ,D_{2222} , D_{3333}, D_{1122},D_{1133},D_{2233}, D_{1112},D_{1113}, D_{1222},D_{2223},
D_{1333}, D_{2333}, D_{1123}, D_{1223}, D_{1233} \bigr),
\end{align*}
\begin{minipage}{\textwidth}
\footnotesize
\noindent
\begin{align*}
 & B = 
 \arraycolsep=1.4pt\def\arraystretch{2.2}
\left(\begin{array}{ccccccccccccccc} 
 \frac{1}{2 } & \frac{1}{2 } & \frac{1}{2 } & \frac{1}{6 } & \frac{1}{6 } & \frac{1}{6 }
 & 0 & 0 & 0 & 0 & 0 & 0 & 0 & 0 & 0 
\\ 0 & 0 & 0 & 0 & 0 & 0 & \frac{\sqrt{15}}{8 } & 0 & \frac{\sqrt{15}}{8 } & 0 & 0 & 0 & 0 & 0 &
 \frac{\sqrt{15}}{24 }
\\ 0 & 0 & 0 & 0 & 0 & 0 & 0 & 0 & 0 & \frac{\sqrt{15}}{8 } & 0 & \frac{\sqrt{15}}{8 } & \frac{\sqrt{15}}{24 } 
& 0 & 0\\ -\frac{\sqrt{5}}{4 } & -\frac{\sqrt{5}}{4 } & \frac{\sqrt{5}}{2 } & -\frac{\sqrt{5}}{12 } & \frac{\sqrt{5}}{24 } & \frac{\sqrt{5}}{24 } 
& 0 & 0 & 0 & 0 & 0 & 0 & 0 & 0 & 0\\ 0 & 0 & 0 & 0 & 0 & 0 & 0 & \frac{\sqrt{15}}{8 } & 0 & 0 & \frac{\sqrt{15}}{8 } & 0 & 0 & \frac{\sqrt{15}}{24 }
 & 0\\ \frac{\sqrt{15}}{2 } & -\frac{\sqrt{15}}{2 } & 0 & 0 & \frac{\sqrt{15}}{12 } & -\frac{\sqrt{15}}{12 } 
& 0 & 0 & 0 & 0 & 0 & 0 & 0 & 0 & 0\\ 0 & 0 & 0 & 0 & 0 & 0 & \frac{3 \sqrt{35}}{16 } & 0 & 0 & -\frac{3 \sqrt{35}}{16 }
 & 0 & 0 & 0 & 0 & 0\\ 0 & 0 & 0 & 0 & 0 & 0 & 0 & 0 & 0 & -\frac{\sqrt{30}}{16 } & 0 & 0 & \frac{\sqrt{30}}{16 } 
& 0 & 0\\ 0 & 0 & 0 & 0 & 0 & 0 & -\frac{3 \sqrt{5}}{16 } & 0 & -\frac{3 \sqrt{5}}{16 } & 0 & 0 & 0 & 0 & 0 & \frac{3 \sqrt{5}}{8 }
\\ 0 & 0 & 0 & 0 & 0 & 0 & 0 & 0 & 0 & -\frac{9 \sqrt{10}}{32 } & 0 & \frac{3 \sqrt{10}}{8 } & 
-\frac{3 \sqrt{10}}{32 } & 0 & 0\\ 
\frac{9}{16 } & \frac{9}{16 } & \frac{3}{2 } & \frac{3}{16 } & -\frac{3}{4 } & 
-\frac{3}{4 } & 0 & 0 & 0 & 0 & 0 & 0 & 0 & 0 & 0\\ 0 & 0 & 0 & 0 & 0 & 0 & 0 & -\frac{9 \sqrt{10}}{32 } & 0 & 0 & \frac{3\sqrt{10}}{8 } & 0 & 0 & -\frac{3 \sqrt{10}}{32 }
 & 0\\ -\frac{3 \sqrt{5}}{8 } & \frac{3 \sqrt{5}}{8 } & 0 & 0 & \frac{3 \sqrt{5}}{8 } & -\frac{3 \sqrt{5}}{8 } & 0 & 0 & 0 & 0 & 0 & 0 & 0
 & 0 & 0\\ 0 & 0 & 0 & 0 & 0 & 0 & 0 & \frac{3 \sqrt{70}}{32 } & 0 & 0 & 0 & 0 & 0 & -\frac{3 \sqrt{70}}{32 } & 0\\ 
\frac{3 \sqrt{35}}{16 } & \frac{3 \sqrt{35}}{16 } & -\frac{9 \sqrt{35}}{8 } & 0 & 0 & 0 & 0 & 0 & 0 & 0 & 0 & 0 & 0 & 0 & 0 
\end{array}\right) 
\frac 1 {\sqrt{\pi}}
\end{align*} 
\end{minipage}
\\

\noindent
Next
we discuss  the prior distribution for the spherical harmonic coefficients.
When these are independent Gaussian random variables with 
\begin{align*}
E\bigl( \theta_{2\ell,m}  \bigr)=0, \quad E\bigl( \theta_{2\ell,m}^2 \bigr)= a_{2\ell}^2,   \quad \ell \in \N, \; - 2\ell \le m \le 2\ell\;,
\end{align*}
it follows from Theorem 6.11  in  (\cite{peccati:marinucci})   that 
$(d(u):u\in S^2)$ is an    isotropic, centered and symmetric  Gaussian random field. Moreover all the
random fields in this class are obtained in such a way,
 and are characterized by their {\it angular power spectrum} $( a_{2\ell}^2, \,\ell \in \N)$.
Consequently the tensor coefficients $(D_{\kappa}: \kappa\in \N^3, |\kappa|=2n)$ are also centered Gaussian random variables with covariance
\begin{align*} \Omega^{-1} =  B^{\top} A B,
\end{align*}
where the diagonal matrix $A$ is the covariance of the spherical harmonic coefficients 
\begin{align*}
( \theta_{2\ell,m},\, 0\le\ell \le n,\, -2\ell\le m \le 2\ell ).
\end{align*}
After inverting the covariance and comparing with the precision matrices $\Omega$  in  (\ref{Omega_D:matrix:2nd}) and (\ref{Omega_D:matrix:4th}), we find 
the following linear correspondances between  precision parameters:
for the 2nd-order tensor model 
\begin{align*}
\eta= \biggl(\frac{  8 \,\pi }{15}  \biggr) a_2^{-2} , \quad
\lambda =\biggl( \frac{ 4 \,\pi } 9  \biggr) a_0^{-2} -  \biggl( \frac{ 8 \, \pi}{45} \biggr)a_2^{-2}, \quad \delta= \biggl(\frac{ 4 \,\pi } 3  \biggr) a_0^{-2} \; ,
\end{align*}
and for the 4th-order tensor model 
\begin{align*} & 
\eta=
\biggl(\frac{48 \,\pi}{245}
\biggr)a_2^{-2} +\biggl( 
\frac{128  \, \pi}{2205}
\biggr) a_4^{-2}, \quad
\lambda = \biggl(
\frac{4 \,\pi}{25 } \biggr) a_0^{-2}  + 
\biggl(\frac{ 16 \, \pi}{ 245 } \biggr) a_2^{-2 }
-\biggl(\frac{ 128 \, \pi }{ 3675} \biggr) a_4^{-2} , & \\ &
\gamma =-\biggl(\frac{48\, \pi}{245} \biggr)
a_2^{-2} +
\biggl(\frac{32\, \pi}{735}\biggr) a_4^{-2}  , & \\ &
\delta= \biggl(\frac{12\, \pi}{5} \biggr)
a_0^{-2}, \quad  \beta=\biggl(\frac{48\, \pi}{35} \biggr) a_2^{-2}, \quad \alpha=
\biggl(\frac{32\, \pi}{315}\biggr)a_4^{-2} \; .
\end{align*}
 When the diffusivity function is assigned voxelwise as  
\begin{align*}
  d_v(u)=\sum_{\ell= 0 }^{n} \sum_{m=-2\ell}^{2\ell} \theta_{2\ell,m}(v) Y_{2\ell,m}(u) , \quad  v\in V, u\in S^2 ,
\end{align*}
with  common truncation level $n$, we define the (improper) regularization prior for the random field
by assigning a Gaussian  prior to the coefficients' pairwise differences as follows
\begin{align*} 
 \pi\bigl( \theta_{2\ell,m }(v) : 0 \le \ell \le n, -2\ell\le m \le 2\ell  \bigr) \propto \prod\limits_{\ell=0}^n  a_{2\ell}^{-(4\ell +1) |V| } 
\exp\biggl(  - \frac 1 2 \sum_{\ell=0}^n a^{-2}_{2\ell} \sum_{m=-2\ell}^{2\ell} \sum_{v \sim w} \bigl\{  
 \theta_{2\ell,m}(v)-\theta_{2\ell,m}(w) \bigr\}^2  \biggr).
\end{align*}
The Bayesian computations of Sections \ref{mcmc:update},\ref{dependence_prior},  apply directly with parameter
\begin{align*}
\theta(v)=\bigl( \log S_v(0), \theta_{2\ell,m}(v) :  0 \le \ell \le n, -2\ell\le m \le 2\ell \bigr)^{\top}\in\R^{1+d} 
,\; d=(2n+1)(n+1),\end{align*}
 design matrix $Z\in\R^{m\times(1+d)}$  with rows 
\begin{align*}
  Z({\bf q})= \bigl( 1, - b  Y_{2\ell,m}( u) : 0 \le \ell \le n, -2\ell \le m \le 2\ell \bigr) , \quad  u = { \bf q} / |{\bf q}| ,\; b= |{\bf q}|^2 /2
\; ,
\end{align*}
and diagonal  precision matrix $\Omega\in \R^{(1+d)\times(1+d)}$ with  diagonal entries
\begin{align*}
\bigl(\rho,a_0^{-2},a_{2}^{-2},a_{2}^{-2},a_{2}^{-2},a_{2}^{-2},a_{2}^{-2} ,\dots,\underbrace{a_{2n}^{-2},\dots,a_{2n}^{-2}}_{\mbox{$(4n+1)$ times}}\bigr) \;.
\end{align*}

Assuming  an improper and scale invariant prior  for the angular power spectrum, given as 
\begin{align*}
 \pi (a_{2\ell,m}^2: 0\le\ell\le n) \propto \prod\limits_{\ell=0}^n  a_{2\ell}^{-2},
\end{align*}
 we obtain   the full conditional distribution for the precision coefficients as
\begin{align*}
 \pi( a_{2\ell}^{-2} |  \theta_{2\ell,m}(v): v \in V , -2\ell \le m \le 2\ell) \sim
\mbox{Gamma}\biggl(  (2\ell +1/2) |V| , \frac  1 2 \sum_{m=-2\ell}^{2\ell} \sum_{v \sim w} \bigl\{  
 \theta_{2\ell,m}(v)-\theta_{2\ell,m}(w) \bigr\}^2  \biggr).
\end{align*}
In the McMC the angular power spectrum is then  updated by sampling independently from these
full conditionals and taking the  inverse.

\newpage
\section{Results }\label{results}
In the follow-up, we illustrate the performance of our method with a real data example.

\paragraph{ The dataset }
The data consists of $4596$ diffusion MR-images of the brain of an  healthy human volunteer,
 taken from four $5mm$-thick consecutive axial slices, and measured with  a
 Philips Achieva $3.0$ Tesla MR-scanner.
The image resolution is
$128\times 128$ pixels with size $1.875\times 1.875$ $mm^2$. 
 After masking  out the skull and the ventricles, 
we remain with a region of interest (ROI) $V$ containing $18764$ voxels.
 In the protocol  we used all the combinations of the
$32$ gradient directions listed in Table \ref{gradient:directions}, 
with the  $b$-values   in Table \ref{bvalues},
  varying in the range $0-14000 s/mm^2$, with $2-3$ repetitions, for a  total of $23323644$ data points.
\paragraph{  McMC implementation }
The data is analyzed under  2nd and 4th-order  tensor models,  with and without Bayesian regularization, estimating  the
regularization parameters in the first case.  In the Markov chain Monte Carlo we do not
 impose positivity constraints on the tensors as we discussed in Section
\ref{sub:posi}, since we want to count the voxels where the posterior expectation of the tensor
is non-positive.
To begin with, we  compute  independently at each voxel $v$ a preliminary estimator
for the tensor  and noise parameters $\theta(v),\sigma^2(v)$, obtaining 
the initial state of the Gibbs-Metropolis Markov chain.
This is done under the log-Gaussian 
approximation  discussed in Section \ref{Rice:section},
 by the method of weighted least-squares, and using only observations in the  low $b$-value range $(b< 5000 \; s/mm^2  \; )$. 
For the regularized model, at each McMC-cycle we divide $V$ into blocks,
where each block is the intersection of $V$ with a ball of radius $r=7$ under the graph distance,
and can contain up to 342 voxels.
Since  blocks are separated by at least one voxel, the parameters from different blocks are conditionally independent
given the exterior boundary values, and it is possible to update the blocks in parallel. 
The centers of the blocks are then cyclically shifted at each McMC cycle, and
 at the end of each cycle we also update the
regularization parameters.
 The Markov chain was running for  $25050$ and $22100$ cycles  respectively, under 2nd and 4th-order tensor models, which
took  $257$ and $225$ CPU hours on a 15-core Intel Xeon  E5-2670  processor.

\begin{figure}[!hbtp] 
 \centering
\begin{subfigure}[b]{\textwidth}
\includegraphics[width=1\textwidth]{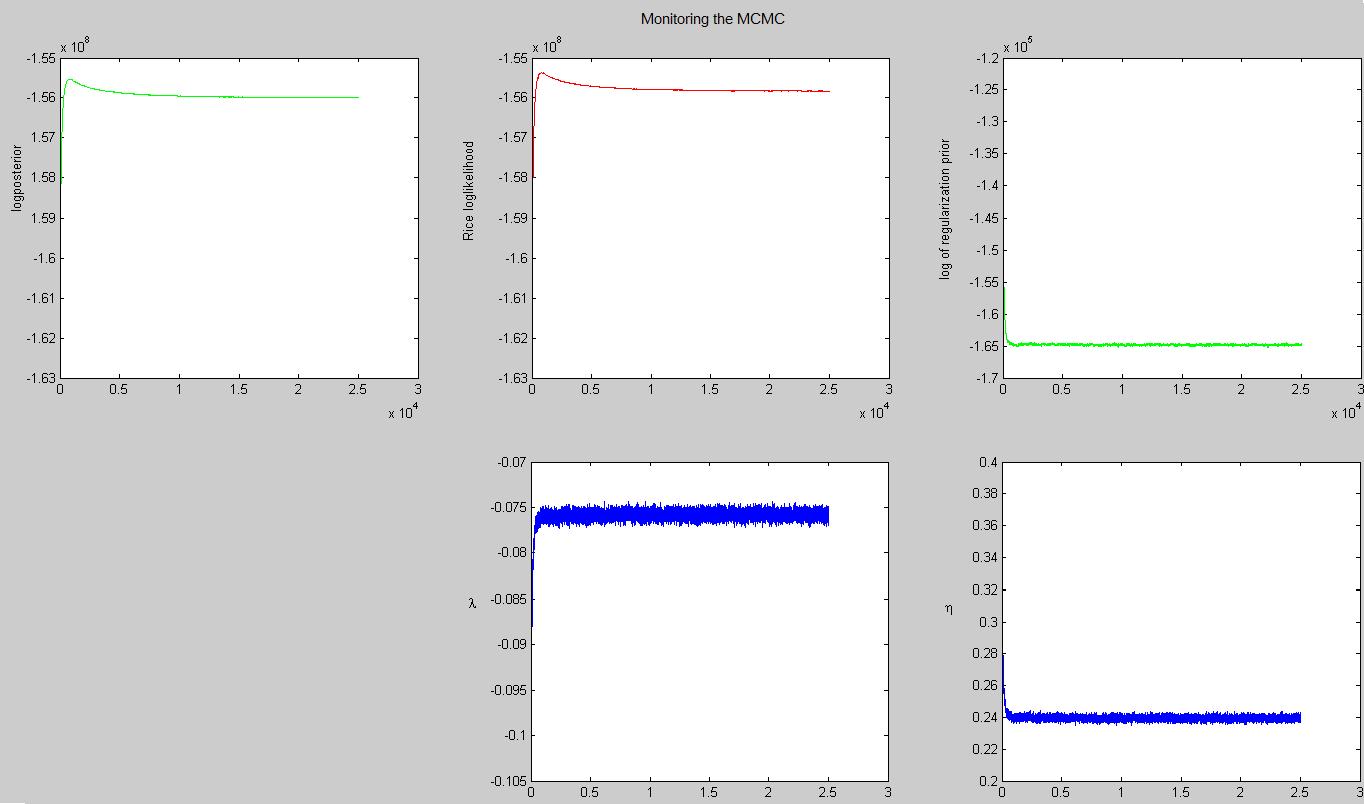}
  \caption{       \label{fig:2ndRad7Cov}               2nd order tensor model, 25050 cycles}
\end{subfigure}
\begin{subfigure}[b]{\textwidth}
\includegraphics[width=1\textwidth]{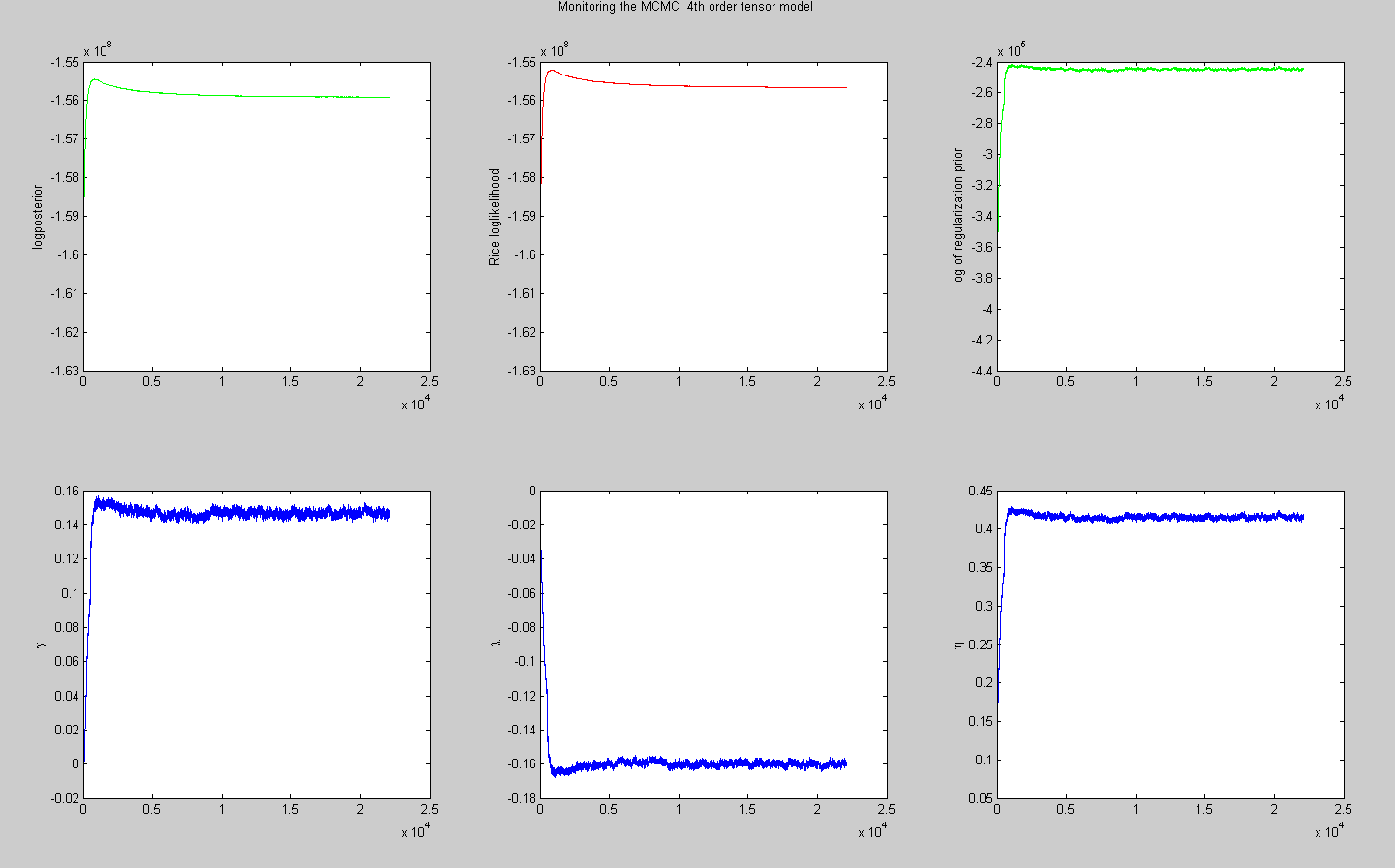}
  \caption{    \label{fig:4thRad7Cov}          4th order tensor model, 22100 cycles}
\end{subfigure}
\caption{ \label{fig:cov} McMC traces of  total posterior density,   likelihood and  prior (in logarithmic scale), 
and regularization parameters $\lambda,\eta$ and $\gamma$, 
for  2nd   
and 4th-order tensor models.        }
\end{figure}

\paragraph{ Monitoring the McMC}
Before computing empirical averages, we waited for the Markov chain to reach stationarity.
The burnin-period  ($15600$ and   $10450$ cycles under the 2nd and 4th-order tensor models,  respectively)
  was selected by monitoring the logposterior and the regularization
parameters of the samples shown in Fig. \ref{fig:cov}, which deserves an explanation. We see that 
 the Rice-loglikelihood increases first very rapidly, and then decreases before stabililizing.
Such phenomena is not uncommon in high dimensional models, when a maximum likelihood estimator
is used to construct the initial configuration (see for example Fig. 3  in \cite{besag}).
To see this effect in a toy model, just
consider a  Gaussian vector  $X\in \R^n$ with  i.i.d. coordinates $X_i\sim{\mathcal N}(\theta,\sigma^2)$, which satisfies
\begin{align} \label{toy:example}
    \sup_{x\in \R^n }  \bigl\{  \log p_n(x) \bigr\} - E_P\bigl(\log p_n(X) \bigr )  = \frac n 2   \; .
\end{align}
In high dimension, under the posterior distribution the typical configuration
and the maximum a posteriori (MAP) configuration can be very different, 
with a set of typical configurations containing most of the probability mass, while the probability
mass concentrated around the MAP-configuration is negligible. 
Since we start the Markov chain from the maximum likelihood estimator under the approximating
log-normal model, at the beginning the orientation of all tensors (but not their eigenvalues)
are close to optimal also 
under the exact Rice likelihood model.  Then the  tensor eigenvalues and noise parameters
move rapidily towards  configurations with highest posterior probability. After this phase,
it takes a while for the tensor orientations to mix-up.
 Since the acceptance probabilities are not uniform betwteen blocks, and we are estimating simultaneously the regularization parameters,
the total logposterior density shows a slow decay before reaching stationarity.

\begin{figure}[!hbtp]
\centering
\begin{subfigure}[b]{0.5\textwidth}
\includegraphics[width=\textwidth]{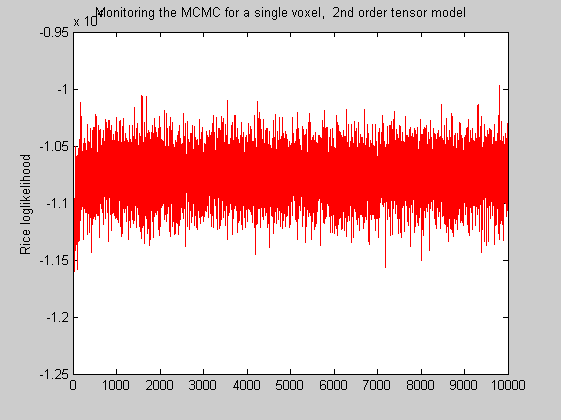}
                \caption{2nd order tensor model, 10000 cycles}
        \end{subfigure}%
 \begin{subfigure}[b]{0.5\textwidth}          
  \includegraphics[width=\textwidth]{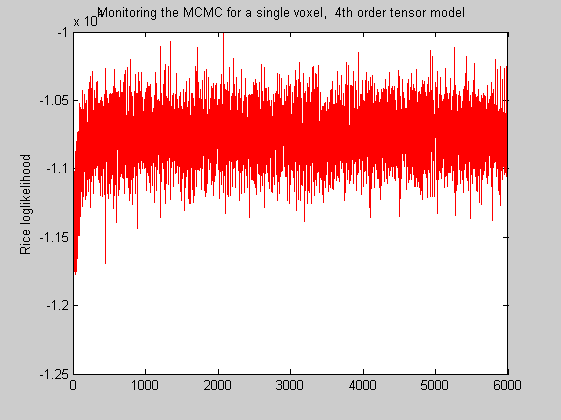}         
          \caption{4th order tensor model, 6000 cycles }
       \end{subfigure}%
    \caption{\label{fig:indep} McMC trace of the Rician loglikelihood for a single voxel, under the  2nd and 4th-order tensor models (without Bayesian regulatization) }
\end{figure}

For comparison, we plot in Fig.  \ref{fig:indep} the McMC trace of the
 Rician loglikelihood for a single voxel under 2nd and 4th tensor models,  
without Bayesian regularization, which  converges rapidly to stationarity.

    
\if
In Fig. \ref{fig:cov}, we can see that all the chains got convergence after certain periods. 
In 2nd order DTE, the logarithmic Rician posterior is getting to stabilize
 roughly after 15,000 iterations (iters.). While, the converging situation
 of the smoothing parameters $\lambda$ and $\eta$ happened only after 5,000 iters.
 In 4th order scenario, all the chains are getting to convergence after 10,000 iters.

In order to get better estimation of the tensor parameters, based on McMC a recognized burnin period should be ignore.
 For 2nd order case, we chose the first 15,600 iters. as the burnin, then compute the average values
 of tensor parameters from the last 9,400 iters. as the tensor estimates. 
The estimates of smoothing parameters ($\lambda$ and $\eta$) can be calculate 
after e.g. 5,000 iter., since they converged much faster than the others. 
For 4th order case, all the parameters of interest go to stabilize after roughly 10,000 as the burnin.
 So we compute the empirical means after the first 10,450 iters..  

In the histories of Rice densities, we got some strange curves (also see \cite{}) at the
 beginning where the traces go up rapidly to a maximum, then go down a little bit to stabilize. 
One main reason leads such situation is the effects of regularization. The other reason is that some voxels
 have very low acceptance rates which effect the monitors of whole voxels. 
Fig.\ref{fig:singleTensor} show the MCMC monitors of two random picked up but good 
(not from artefacts) voxels without regularization, where we run 10,000 and 6000 independent updated
 with 1000 burnin for 2nd and 4th order tensor model, respectively. 
\fi

\paragraph{Acceptance probabilities}

\begin{figure}[!hbtp]
\centering
\begin{subfigure}[b]{0.492\textwidth}
\includegraphics[width=\textwidth]{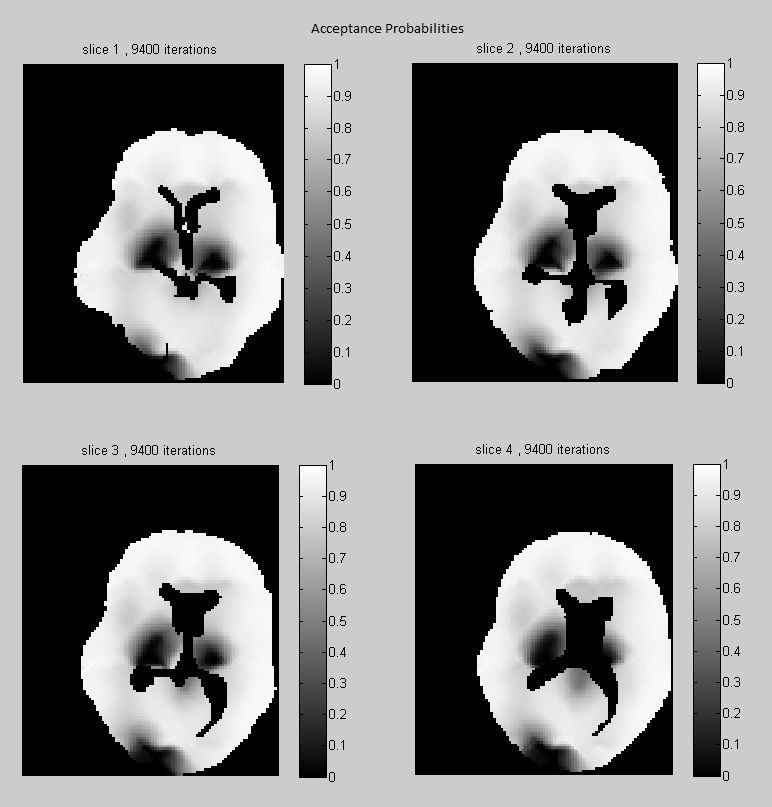}
                \caption{\label{fig:AVACPT2} acceptance probability, 2nd order tensor model}
        \end{subfigure}%
 \begin{subfigure}[b]{0.5\textwidth}
                \includegraphics[width=\textwidth]{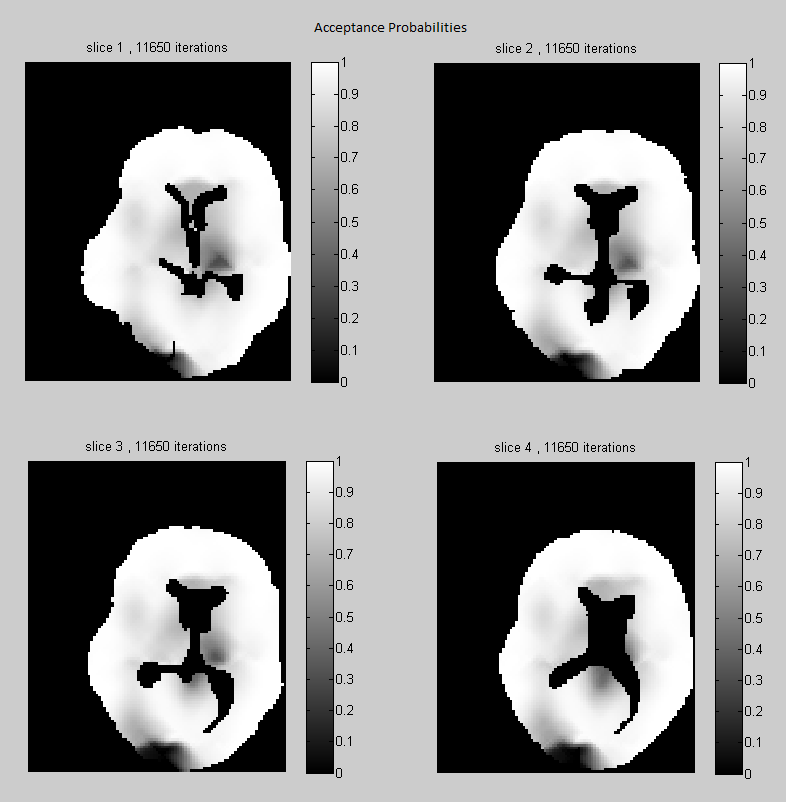}
                \caption{\label{fig:AVACPT4}acceptance probability, 4th order tensor model}
        \end{subfigure}         
   \caption{ \label{fig:accept_regularized} Acceptance probabilities in grey level scale (black=0,white=1) 
for the 2nd and 4th-order regularized tensor models } 
\end{figure}

In Fig. \ref{fig:accept_regularized} we show the acceptance probabilities for the Gibbs-Metropolis block update of the
tensor parameters, estimated for each voxel 
under the regularized 2nd and 4th order tensor models. Note  that, although we use large block updates with more than
300 voxels in each block, the acceptance probabilities are remarkably high in most of the voxels (see the histograms). 
It means that in most cases the our Gaussian approximation is very close to the exact full conditional distribution of the
tensor parameters in a block. Note also that in Fig. \ref{fig:AVACPT2} (which corresponds to 2nd order tensor model)
 there are some regions with lower acceptance probability. In such areas one should use update blocks of smaller size.
These regions of  lower acceptance probability are either artefacts, where the data are corrupted,
or contain complex structures where the 2nd order tensor model does not fit well the data, and a higher
order model would be more appropriate. We see two low acceptance probability regions situated symmetrically
on the left and right sides of the ventricles. Anatomically this corresponds to the corona radiata where
fiber bundles from multiple directions are crossing. By comparing with   Fig. \ref{fig:AVACPT4}
we see that in these regions  the acceptance probability  improves under the (regularized) 4th order tensor model.
For the  diffusion model without regularization, the independent tensor updates have
high acceptance probabilities   at all voxels, under both 2nd and 4th-order tensor models (in \ref{fig:hist_acc_indep}).



\begin{figure}[!hbtp]
\centering
\includegraphics[width=0.85\textwidth]{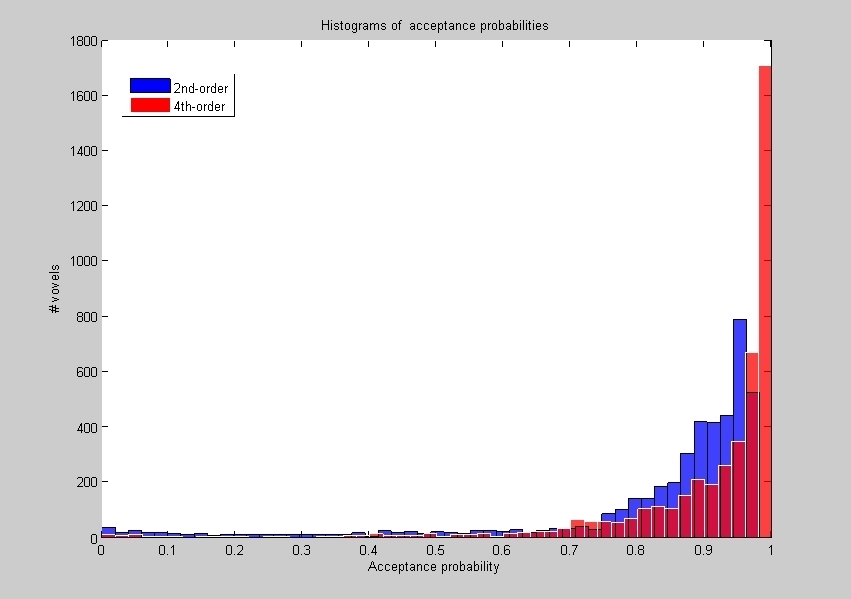}
   \caption{ \label{fig:acc_hist}Acceptance probabilities across voxels for tensor block updates, 
 under 2nd  and 4th order regularized tensor  models. }
\end{figure}

\if
\begin{figure}[!hbtp]
\centering
\begin{subfigure}[b]{0.492\textwidth}
\includegraphics[width=\textwidth]{indep_acceptance_2}
                \caption{  \label{fig:acc_indep_2} acceptance probability, 2nd order tensor model}
        \end{subfigure}%
 \begin{subfigure}[b]{0.5\textwidth}
                \includegraphics[width=\textwidth]{indep_acceptance_4}
                \caption{  \label{fig:acc_indep_4}acceptance probability, 4th order tensor model}
        \end{subfigure}         
   \caption{  \label{fig:acc_indep_2_and_4} Acceptance probabilities 
for the 2nd and 4th-order  tensors, updated independently, without regularization  }
\end{figure}
\fi

\begin{figure}[!hbtp]
\centering
\includegraphics[width=1\textwidth]{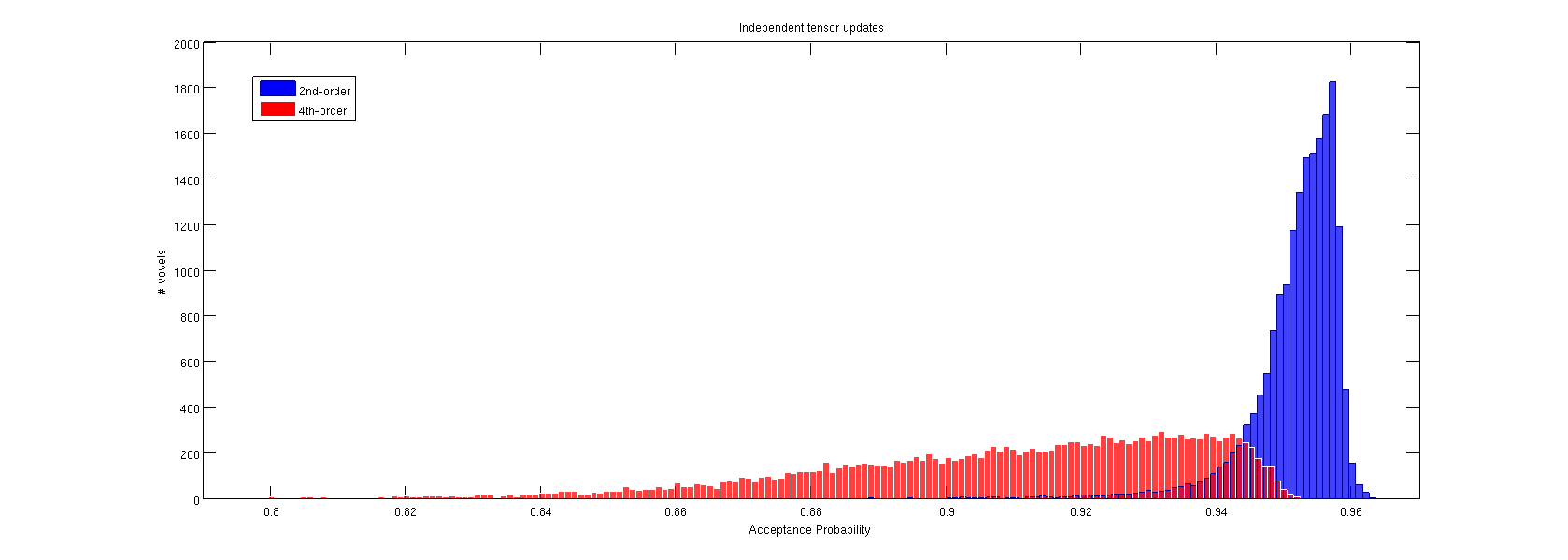}
   \caption{ \label{fig:hist_acc_indep}Acceptance probabilities across voxels for tensor independent updates, without regularization,
 under 2nd  and 4th order models.}
\end{figure}
\begin{figure}[!hbtp]
\centering
\begin{subfigure}[b]{0.45\textwidth}
\includegraphics[width=\textwidth]{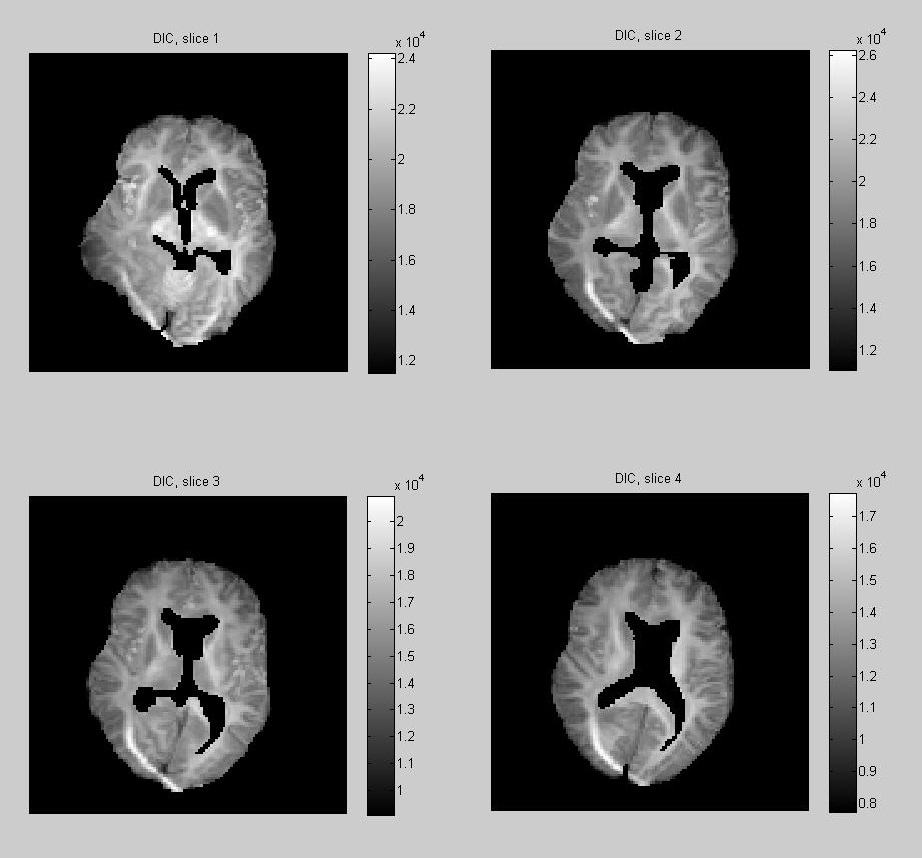}
                \caption{   \label{fig:DIC2}2nd order independent tensor model}
        \end{subfigure}%
 \begin{subfigure}[b]{0.5\textwidth}
                \includegraphics[width=\textwidth]{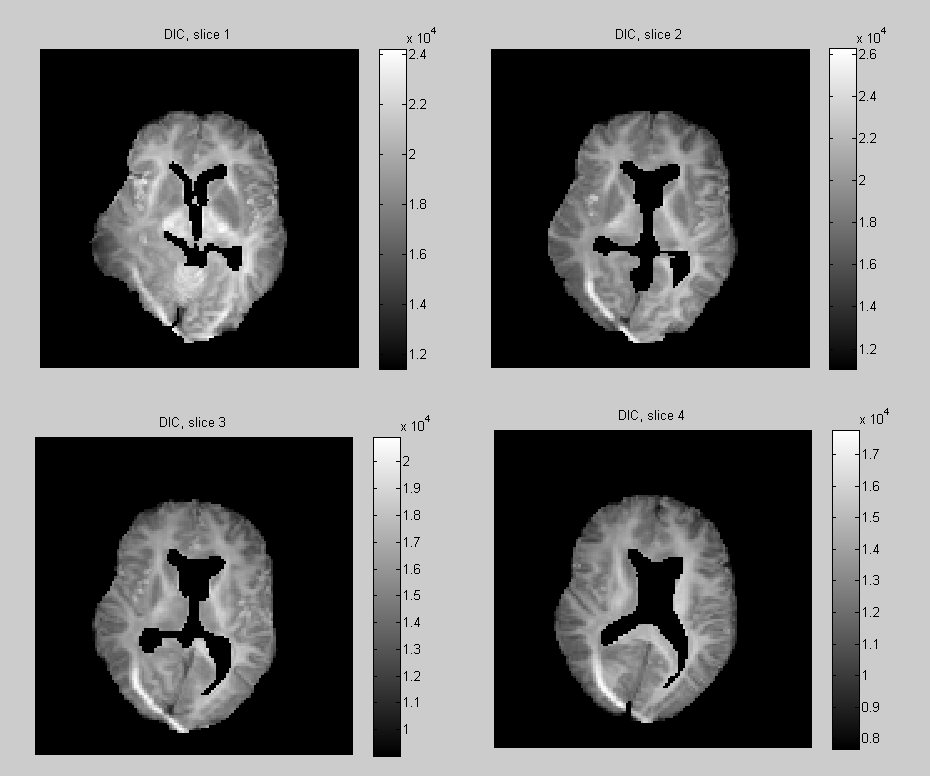}
                \caption{ \label{fig:DIC4}4th order independent tensor model}
        \end{subfigure}         
   \caption{\label{DIC:independent}DIC maps under 2nd 
 and 4th-order tensor model without regularization.
Lower values (dark) correspond to better model fit.} 
\end{figure}
\begin{figure}[!hbtp]
\centering
\begin{subfigure}[b]{0.5\textwidth}
\includegraphics[width=\textwidth]{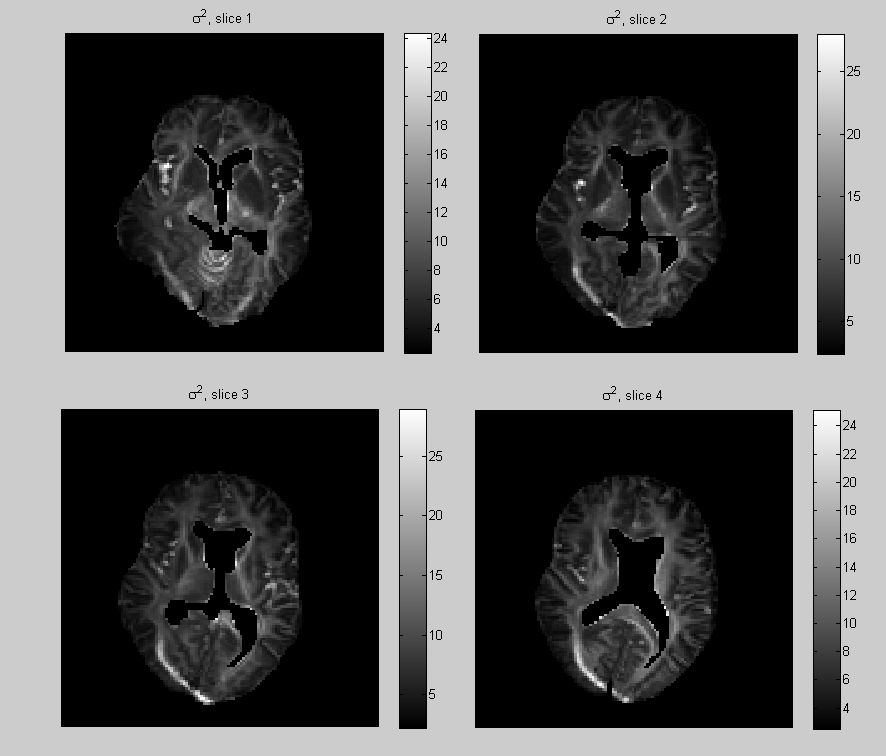}
                \caption{    \label{fig:noise2}posterior expectation, 2nd order tensor model}
        \end{subfigure}%
 \begin{subfigure}[b]{0.43\textwidth}
                \includegraphics[width=\textwidth]{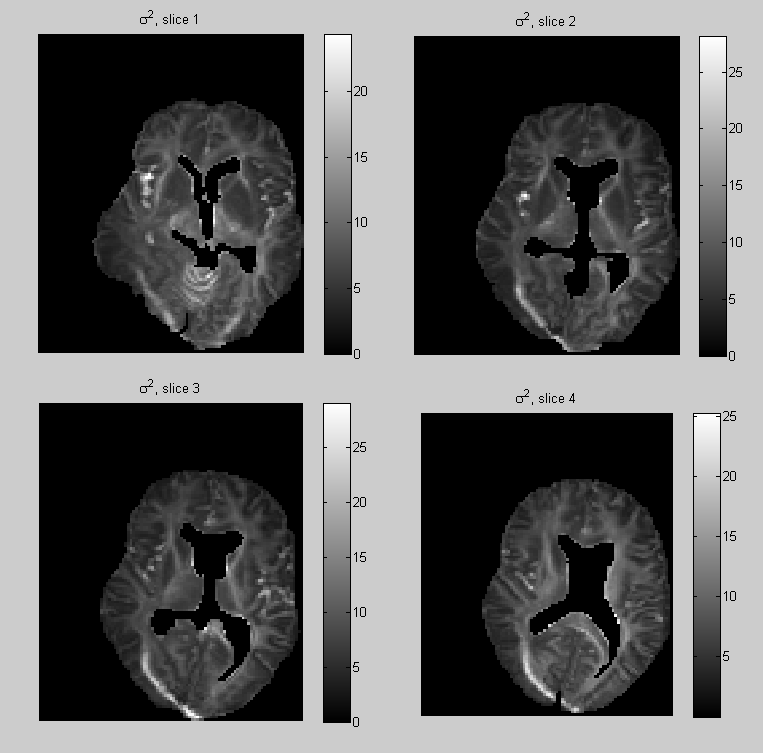}
                \caption{ \label{fig:noise4}posterior expectation 4th order tensor model}
        \end{subfigure}         
   \caption{\label{fig:posterior_noise} Posterior expectations of the variance parameters in the Rician noise distribution, in 2nd and 4th-order tensor models }
\end{figure}

\paragraph{Deviance Information Criterion}
The deviance information criterion (DIC), introduced by  \cite{spiegelhalter}, is a  measure of model fitting used in Bayesian model selection 
as an alternative to
 Bayes factors.  Unlike Bayes factors,  DIC is well defined also when improper priors are assumed, as it is the case in our settings.
It  is defined as
\begin{align*}
   \mbox{DIC} =  2 E_{\pi} \bigl(  D(\theta) \big\vert  \mbox{data} \bigr)-D\bigl(  E_{\pi}(\theta|\mbox{data} ) \bigr) , 
\end{align*} where
$D(\theta) = - 2\log p(\mbox{data} | \theta )$ is the deviance,
and we take conditional expectations with respect to the posterior distribution of the parameters $\theta$.
Defined in analogy with the toy example of  Eq. (\ref{toy:example}), the {\it effective number of parameters }
\begin{align*}
n_{eff}:= D\bigl(  E_{\pi}(\theta|\mbox{data} )\bigr) - E_{\pi}\bigl(  D(\theta) \big\vert  \mbox{data} \bigr)
 \end{align*}
appears as  penalization term in the expression
\begin{align*}
 \mbox{DIC} = -E_{\pi}\bigl( \log p( \mbox{data} |\theta ) \big\vert  \mbox{data} \bigr) + n_{eff} \;.
 \end{align*}
This allows for model comparisons,  lower  DIC meaning a better  fit to the data relatively to the
effective number of parameters.
In Fig. \ref{DIC:independent} the  DIC is computed independently at each voxel
 under the 2nd and 4th-order tensor models (without regularization). Note
that the voxels with the highest DIC corresponds  to artefacts where the data is corrupted,
and the area of high DIC correspond to complex white matter structures.
 We also calculated the overall DIC for all voxel under the model 2nd and 4th-order tensor models with regularization.
The respective values DIC$= -1.5554 \times 10^8 $ and DIC$= -1.5525 \times 10^8$,
 indicate that when we  penalize the model by the effective number of parameters, overall
the 2th-order tensor model fits our data better  than the 4th-order model.
In Fig. \ref{fig:posterior_noise}  the posterior expectation of the noise parameters $\sigma^2(v)$, are shown.
When these are interpreted as residual variances in model fitting, we see that they are consistent with
the DIC.


\if
is a statistical criteria which is widely used  in Bayesian model selection, and  particularly for those posteriors 
obtaining by Markov chain Monte Carlo (MCMC) simulation. It is defined as
\begin{eqnarray}
DIC_m = 2 \overline{D(\theta_m,m)} - D(\overline{\theta_m}, m),
\end{eqnarray}
where the deviance is
\begin{equation}
 D(\theta_m,m) = -2 \log(p(y|\theta_m, m)).
\end{equation}

And $\overline{D(\theta_m,m)} = \mathbb{E}^{\theta_m}[D(\theta_m,m)]$ is the posterior expectation of the deviance
 which measures how well the model $m$ fits the data. 
The idea is that the smaller DIC is, the better the model should be.
\fi 


\if
\begin{figure}[!hbtp]
\centering
\includegraphics[width=\textwidth]{DIC_histogram}
   \caption{ Histograms of DIC across voxels, for 2nd  and 4th order tensor models. We see in both cases a bimodal distribution,
since the design of the experiment contains either 2 or 3 repetitions (see Table \ref{bvalues}).}
\label{DIC:histograms}
\end{figure}
\fi

\if
We also calculated the DIC of all voxels with regularization. 
For 2nd and 4th order tensor model with $radius = 7 $ we have $DIC =  -1.5554e+008$ and $DIC = -1.5525e+008$, respectively. 
Therefore, the DIC criteria shows that the 4th tensor order model is better than the 2nd one.
\fi

\paragraph{Diffusivity profiles}

\begin{figure}[!hbtp] 
 \begin{center}
 \begin{subfigure}[b]{1\textwidth}
                \includegraphics[width=.35\textwidth]{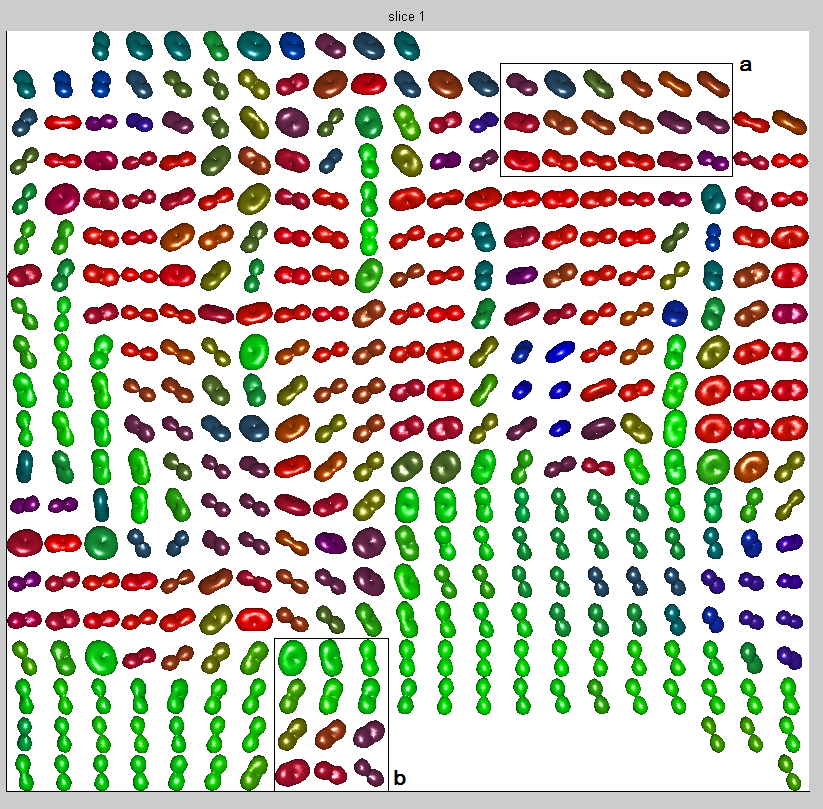}
\includegraphics[width=.6\textwidth]{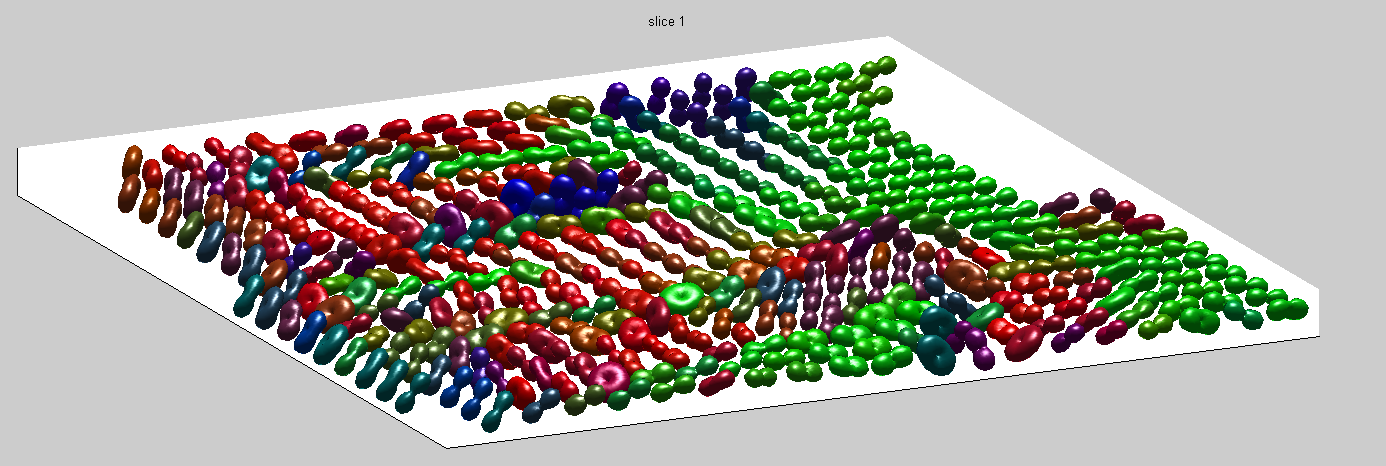}
                \caption{   \label{fig:dti2}Estimated diffusivity profiles   under 2nd-order tensor model}
        \end{subfigure} 
\begin{subfigure}[b]{1\textwidth}
               \includegraphics[width=.35\textwidth]{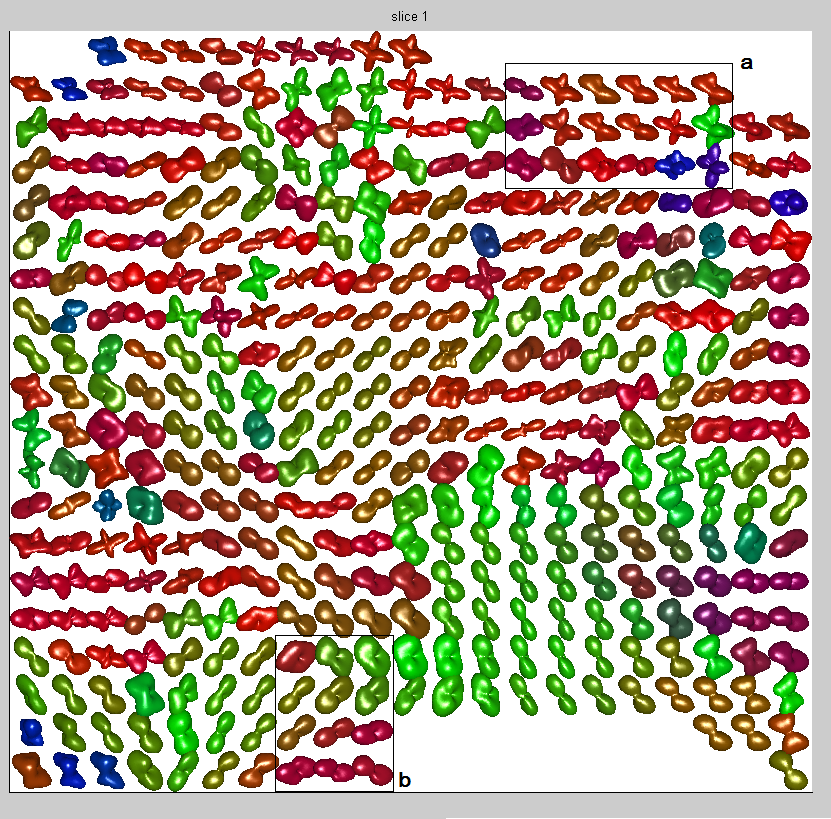}
\includegraphics[width=.6\textwidth]{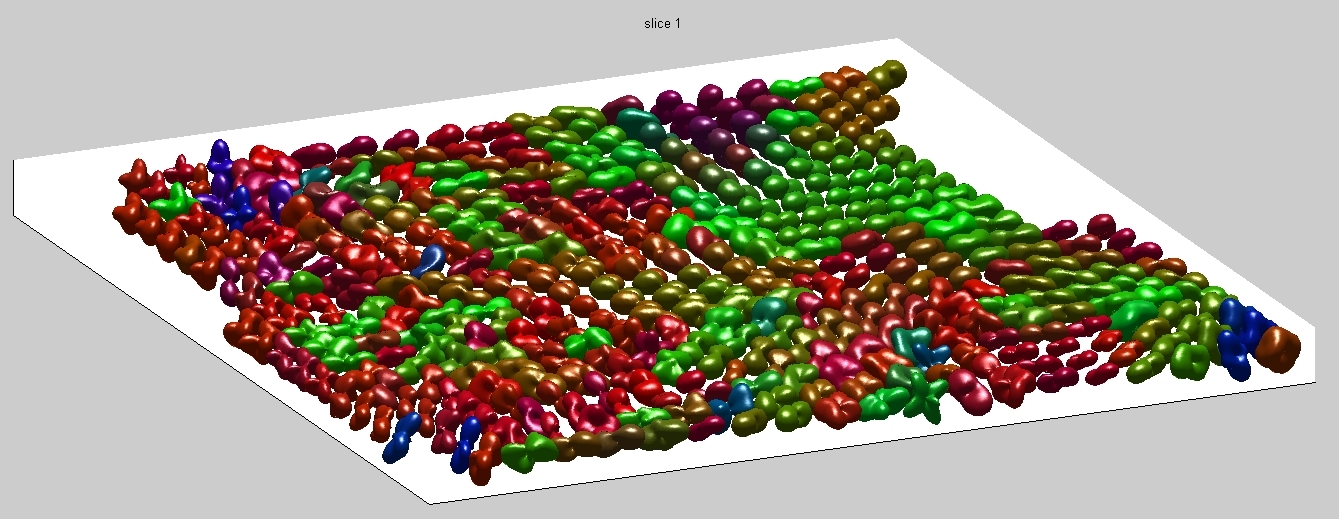}
                \caption{  \label{fig:dti4} Estimated diffusivity profiles   under 4th-order tensor model        }
        \end{subfigure} 
 \end{center}
 \caption{\label{fig:2ja4dtiS1} Estimated diffusivity profiles from a ROI, under 2nd and 4th-order tensor model.
The color-code represents the main direction of the principal eigenvalue of the 2nd-order tensor:
 Red, left-right; Green, anterior-posterior;  Blue, superior-inferior. 
These figures are drawn with the Matlab package fanDTasia written by  Barmpoutis  \citep{barmpoutis,barmpoutis2}. } 
 \end{figure}

\begin{figure}[!hbtp] 
 \begin{center}
 \begin{subfigure}[b]{.43\textwidth}        
\includegraphics[width=\textwidth]{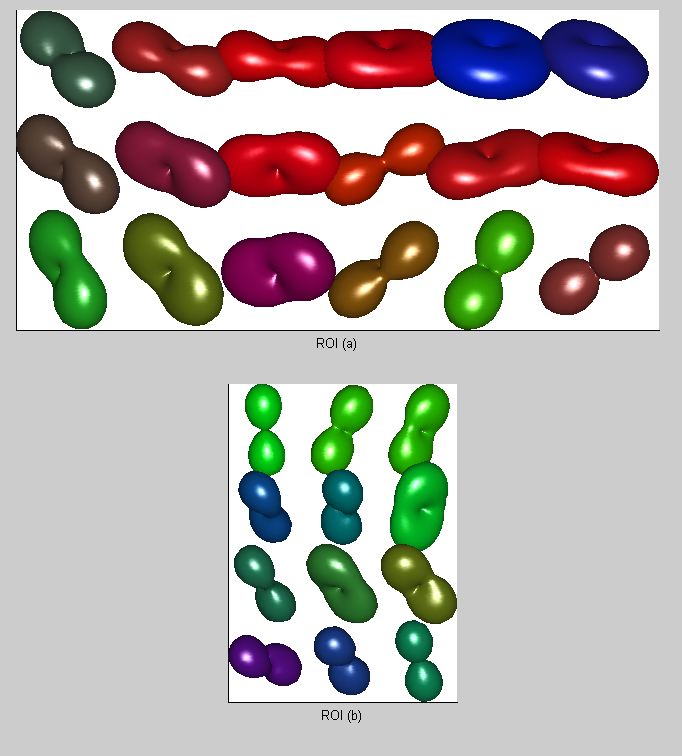} 
               \caption{    \label{fig:DT2} 2nd order   }
        \end{subfigure} 
\begin{subfigure}[b]{.45\textwidth} 
               \includegraphics[width=\textwidth]{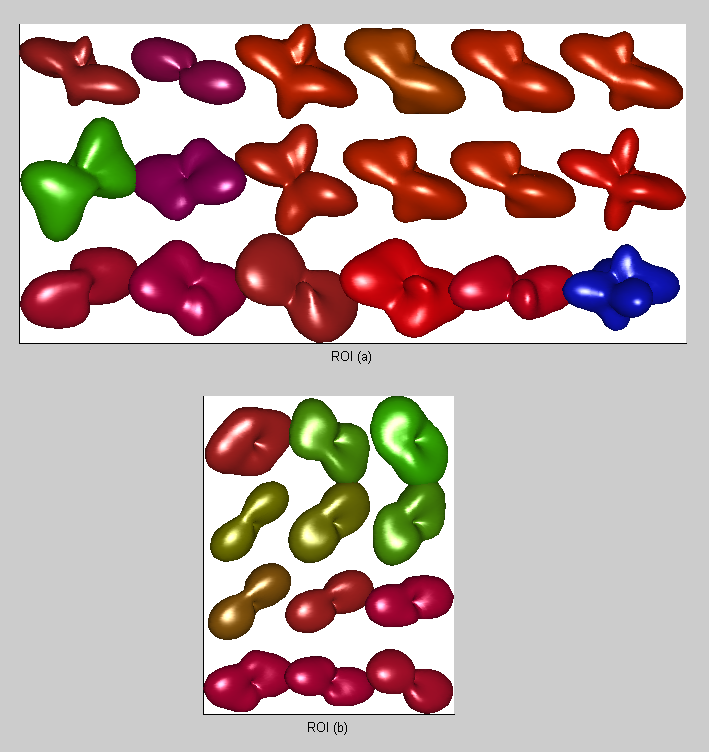}
                 \caption{    \label{fig:DT4}4th-order  }
        \end{subfigure}  
 \end{center}
 \caption{\label{fig:2ja4rank}Estimated diffusivity profiles under 2nd and 4th-order  tensor models in ROI  (a),  showing crossing fibers between the corticospinal
 tract and superior longitudinal fibers, and ROI (b), showing 
fiber crossing near the corpus callosum, both selected from Fig. \ref{fig:2ja4dtiS1}  } 
 \end{figure}

 Fig. \ref{fig:2ja4dtiS1} shows  the diffusivity profiles based on the posterior estimates of the tensors at all voxels in a region of interest.
 For each direction $u\in {\mathcal S^2}$ and spatial location $v\in V\subset \R^3$, 
we plot  the point $( v+\overline{ d_v (u) }  u )\in \R^3$, where $\overline{d_v(u)}$ is the posterior expectation of the diffusivity.
In order to observe the differences between 2nd and 4th order tensor models, in Fig. \ref{fig:2ja4rank}
we zoom into the ROI (a) and (b),
and see that the 4th order tensor model captures the fiber-crossings which  
  the 2nd order model cannot capture. At the  fiber-crossing locations, under the 2nd-order model the two largest 
eigenvalues of the estimated tensor have similar sizes, with a donut-shaped diffusivity profile.

\if
Furthermore, ROI (a) shows crossing fibers between the corticospinal tract and superior longitudinal fibers, and ROI (b) shows 
fiber crossing near the corpus callosum. 
It can be seen that the 4th order angular resolution of tensors provide more detailed information of the water diffusion, 
especially in those complicated fiber-tissue areas.
\newpage
 \fi

\paragraph{Bayesian regularization} 
In Fig. \ref{fig:4rank} we compare  diffusivity profiles from a region of interest  without and with regularization, under the 4th order tensor model.
With regularization, the differences in shape and direction between neighbouring tensors get smoothed. This also
implies noise reduction: the tensor information from data corrupted by  artefacts is corrected  by the information from the neighbours. 
For the 2nd-order tensor model, the regularization effect in the same region was not that evident. 
Since the regularization parameters are not fixed  
but estimated from the data,  we cannot always expect an increase from the
 smoothness  level determined by the data.  In order to achieve a prespecified level of smoothness  we should either fix the regularization
parameters or assign them  a strongly informative prior. The posterior mean and standard deviation of the regularization parameters is given
 in Table \ref{table:post}.
\\

\if
\begin{minipage}{\linewidth}
 \centering 
\label{table:post}
\begin{tabular}{ |l|l |l |l| }
\hline  
   & $\bar\eta$ &  $\bar\lambda$ &  $\bar\gamma$  \\ \hline 
   2nd order&   0.2394 &  -0.0758 &    \\ \hline
   4th  order &  0.4155 & -0.1600  & 0.1469 \\ \hline
\end{tabular}\captionof{table}{Posterior means of regularization parameters}
\end{minipage}
\fi

\begin{minipage}{\linewidth}\renewcommand{\arraystretch}{1.5}
 \centering 
\begin{tabular}{ |l||l |l ||l| l||l|l|} 
\hline    & $\bar\eta$ & $\sqrt{ \overline{\eta^2} -(\bar\eta)^² }$ & $ \bar\lambda$  &   $\sqrt{ \overline{\lambda^2} -(\bar\lambda)^2 }$          
&  $\bar\gamma$  &      $ \sqrt{ \overline{\gamma^2} -(\bar\gamma)^2 } $   \\  \hline \renewcommand{\arraystretch}{1.0}
   2nd order&   0.2394 & 0.0012 & -0.0758 &   $3.9352\times 10^{-4}$   &   &  \\ \hline
   4th  order &  0.4155 &0.0021 & -0.1600  &  0.0012 &         0.1469 &   0.0016 \\ \hline
\end{tabular}\captionof{table}{\label{table:post} Posterior mean and standard deviation of regularization parameters}
\end{minipage}

\if 
\begin{figure}[!hbtp]
\centering
\includegraphics[width=\textwidth]{2nd_order_diffusivity_slice}
\caption{Diffusivity profiles from a ROI under 2nd-order tensor model, estimated with
and without regularization.}
\label{fig:regularization2}
\end{figure}
\fi

\begin{figure}[!hbtp]
\centering
\includegraphics[width=\textwidth]{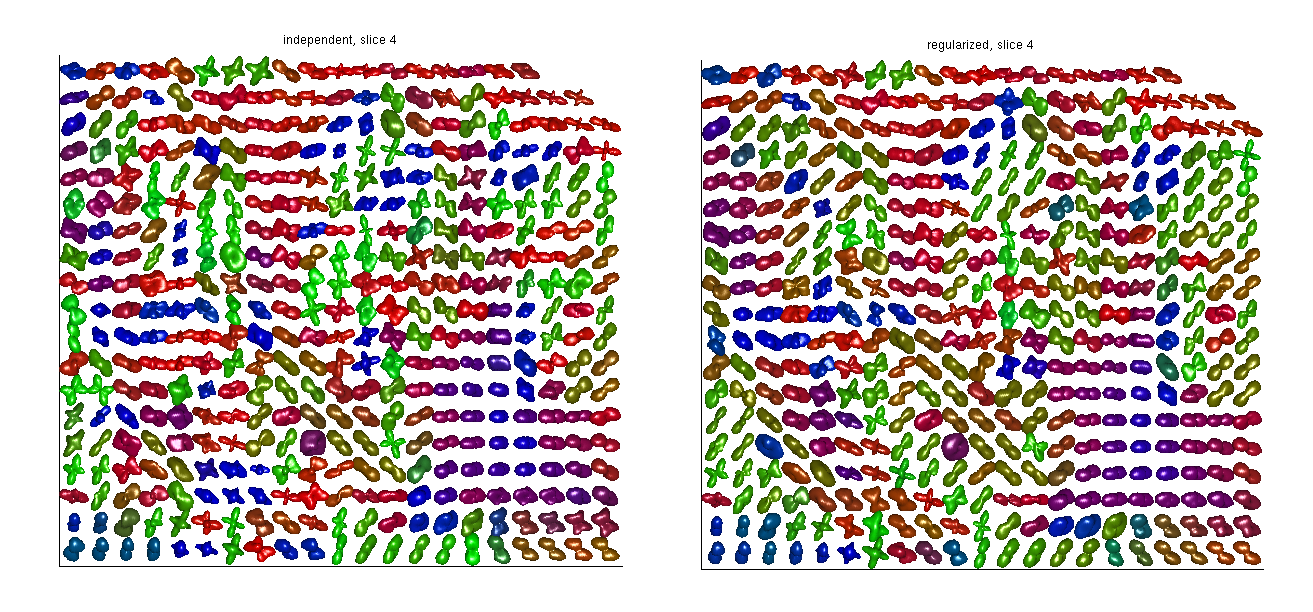}
\caption{\label{fig:4rank}Diffusivity profiles from a ROI under 4th-order tensor model, estimated with
and without regularization.}
\end{figure}
\if
\begin{figure}[!hbtp] 
 \begin{center}
 \begin{subfigure}[b]{.45\textwidth}        
\includegraphics[width=\textwidth]{2ndrad7IndS4} 
               \caption{\label{fig:DT2oS}} 
        \end{subfigure} 
\begin{subfigure}[b]{.45\textwidth} 
               \includegraphics[width=\textwidth]{2ndrad7S4}
    \caption{\label{fig:DT2wS}}  
        \end{subfigure}  
 \end{center}
 \caption{\label{fig:regularization2}Diffusivity profiles from a ROI under 2nd-order tensor model, estimated
with (Fig. \ref{fig:DT2wS})
and  without regularization (Fig. \ref{fig:DT2oS}) 
  } 
 \end{figure}
 \begin{figure}[!hbtp] 
 \begin{center}
 \begin{subfigure}[b]{.45\textwidth}        
\includegraphics[width=\textwidth]{Indd16S4} 
 \caption{\label{fig:DT4oS}}             
        \end{subfigure} 
\begin{subfigure}[b]{.45\textwidth} 
\includegraphics[width=\textwidth]{4thrad7dti}
 \caption{\label{fig:DT4wS}}
        \end{subfigure}  
 \end{center}
\caption{\label{fig:4rank}Diffusivity profiles from a ROI under 4th-order tensor model, estimated with  (Fig.\ref{fig:DT4wS}) 
and without regularization (Fig.\ref{fig:DT4oS}).}
 \end{figure}
\fi

\begin{figure}[!hbtp]
\begin{center}
\begin{subfigure}[b]{0.45\textwidth}
\includegraphics[width=\textwidth]{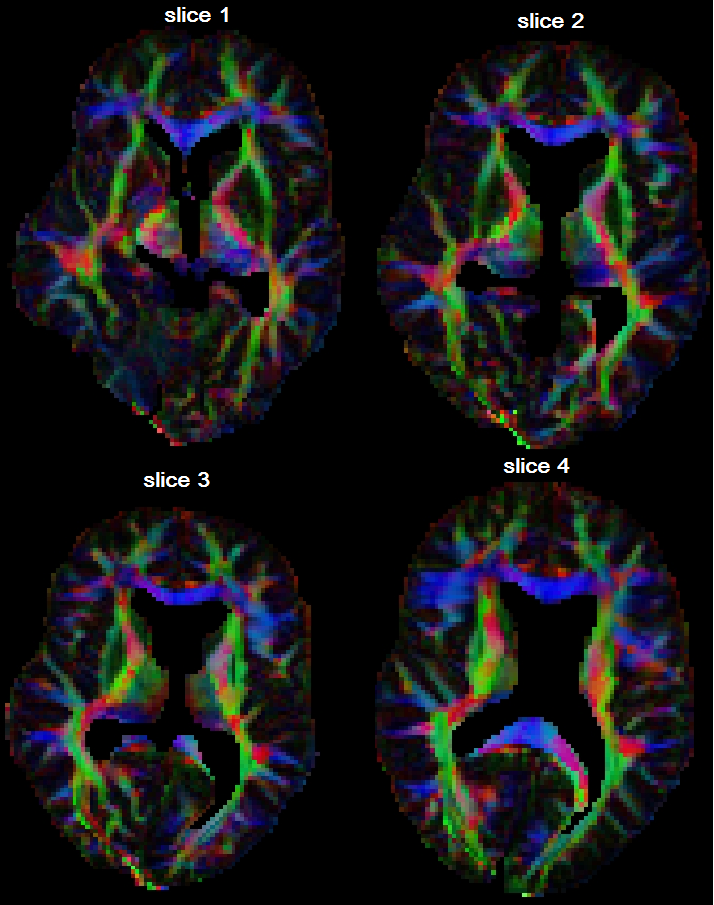}
  \caption{\label{fig:FEFA2}}          
        \end{subfigure}%
 \begin{subfigure}[b]{0.49\textwidth}             
                \includegraphics[width=\textwidth]{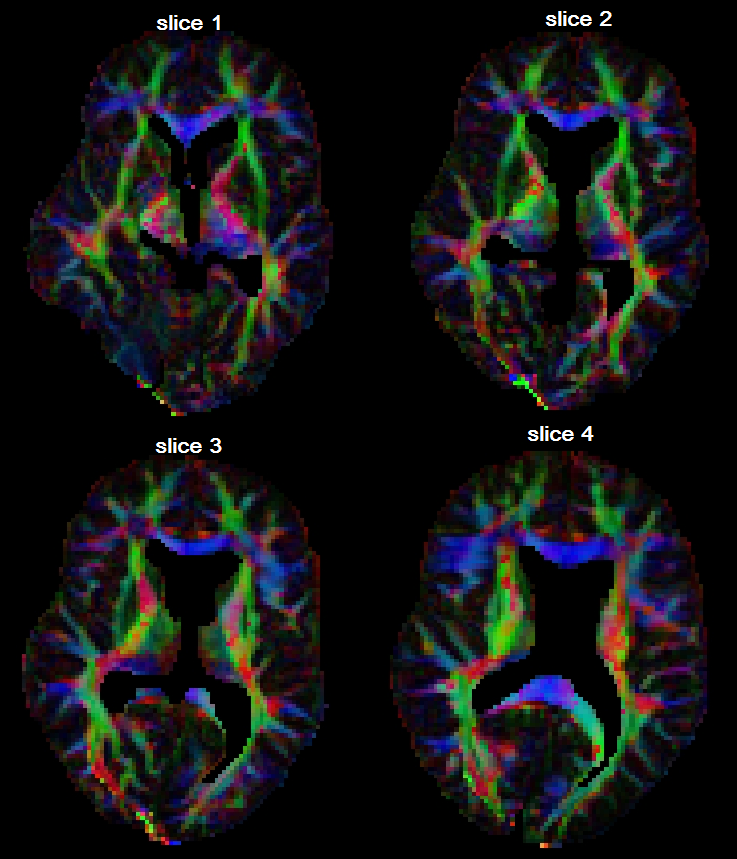}
 \caption{\label{fig:FEFA4thorder}}                
        \end{subfigure}
\end{center} 
   \caption{\label{fig:FA} Bayesian FA estimates  under 
 2nd (Fig. \ref{fig:FEFA2}) and 4th (Fig. \ref{fig:FEFA4thorder}) order tensor models. As in the previous figures,
the color-code  shows the orientations of the principal eigenvalue of the 2nd order tensor, with intensities proportional to 
the fractional anisotropy. }
\end{figure}

\begin{figure}[!hbtp]
\begin{center}
\begin{subfigure}[b]{0.47\textwidth}
\includegraphics[width=\textwidth]{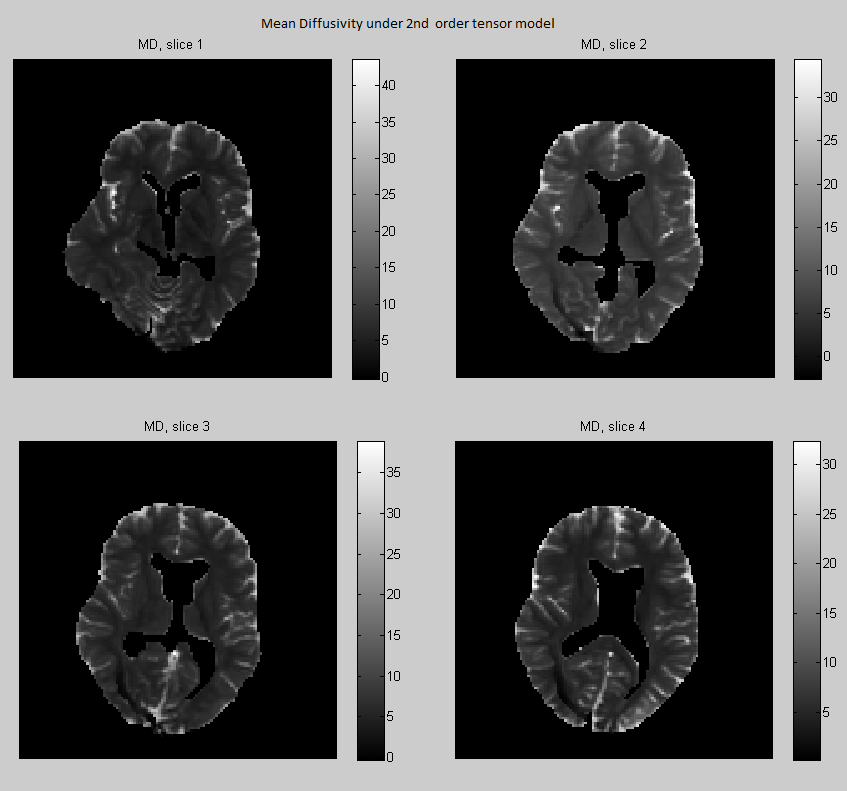}
                \caption{\label{fig:MD2}}
        \end{subfigure}%
 \begin{subfigure}[b]{0.47\textwidth}             
                \includegraphics[width=\textwidth]{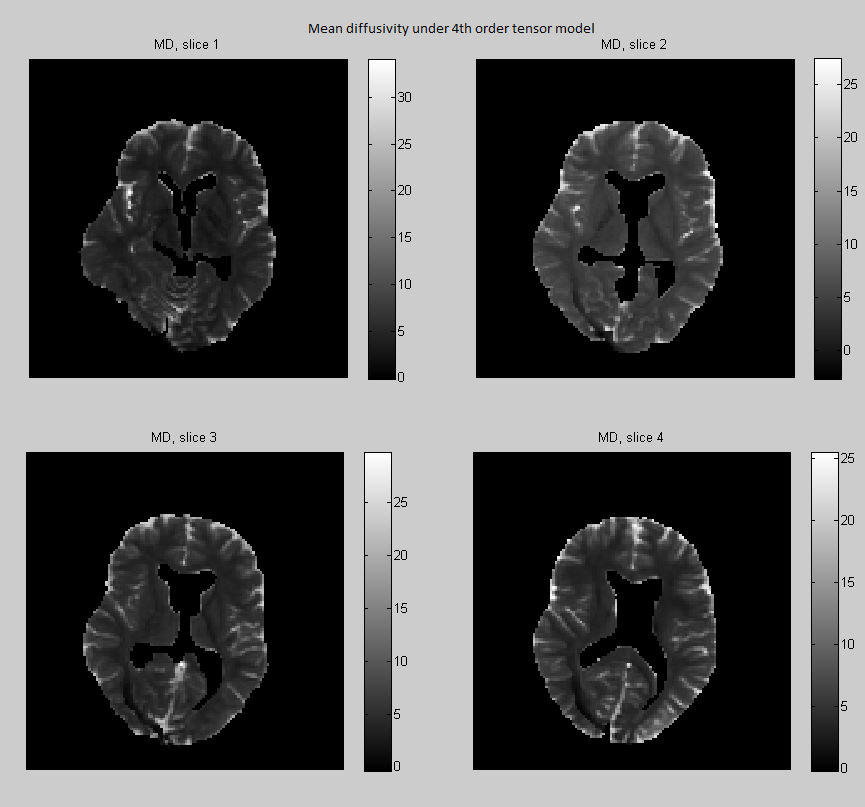}
                \caption{\label{fig:MD4}}
        \end{subfigure}
\end{center} 
   \caption{\label{fig:MD}The mean diffusivity (MD) maps from the results for both 2nd (Fig. \ref{fig:MD2}) and 4th (Fig. \ref{fig:MD4}) order diffusion tensor.}
\end{figure}

\paragraph{Fractional Anisotropy and Mean Diffusivity.}
Fractional anisotropy (FA)
  measures  the degree of anisotropy, while mean diffusivity (MD) is the  average of the diffusivity $d(u)$ function over the unit sphere.
 Both measures are used as biomarkers to study brain pathologies.
These quantities are expressed in terms of the eigenvalues of the 2nd order tensor as
\begin{align*}
    \text{MD}=(\lambda _{1}+\lambda _{2}+\lambda _{3})/3,
\quad \text{FA} ={\frac {{\sqrt {3((\lambda _{1}-MD)^{2}+(\lambda _{2}-MD )^{2}+(\lambda _{3}-MD)^{2})}}}{{\sqrt {2(\lambda _{1}^{2}+\lambda _{2}^{2}+\lambda _{3}^{2})}}}} ,
\end{align*}
In Section  \ref{SH:subsection} we have seen that there is a linear bijection between the tensor coefficients and the coefficients 
of the truncated spherical harmonic expansion of the diffusivity.
  This implies that we can map linearly a 4th-order tensor 
 to a 2nd-order tensor as follows (see \cite{ozars}):
\begin{align*}
D_{11}  &= \frac{3}{35}(
 	9 D_{1111} + 8 D_{1122} + 8 D_{1133} - D_{2222} - D_{3333} - 2 D_{2233})\\
D_{22} &= \frac{3}{35}(
9 D_{2222} + 8 D_{1122} + 8 D_{2233}  - D_{1111}- D_{3333} - 2 D_{1133})\\
D_{33} &= \frac{3}{35}(
9 D_{3333} + 8 D_{1133} + 8 D_{2233}  - D_{1111}- D_{2222} - 2 D_{1122})\\
D_{12} &= \frac{6}{7}(
D_{1112}  + D_{2223} + D_{1233})\\
D_{13} &= \frac{6}{7}(
D_{1113} +D_{1333} + D_{1223})\\
D_{23} &= \frac{6}{7}(
D_{2223} + D_{2333} + D_{1123}) ,
\end{align*}
and
the mean diffusivity  can be also expressed  in terms of the  4th order tensor coefficients as
  \begin{equation}\label{md4}
\text{MD} = \frac{1}{5} (D_{1111} + D_{1122} + D_{1133} + 2 D_{2222} + 2 D_{3333} + 2 D_{2233} =\frac{1}{5} \text{trace}(\widehat{D}),
\end{equation}  
where $\widehat{D}$ was defined in Eq. (\ref{Dhat}).
In Fig. \ref{fig:FA} and \ref{fig:MD} we compare the respectively the Bayesian estimates of FA and MD 
 derived under the 2nd and 4th-order tensor models.


\newpage
\section{Conclusion}
\label{conclusion}
 Rician noise, which models the magnitude of a real valued signal perturbed by additive complex Gaussian noise, 
appears in a  wide range of  applications  in statistics and signal processing.
 By using a novel representation of  the  Rician likelihood, 
we  are able to  reduce  nonlinear regression problems 
with Rician noise to  Generalized Linear Models with Poissonian noise. 
This representation turns out to be very useful in Diffusion
Tensor Imaging, where the problem is to estimate the transition distribution 
of water molecules diffusing inside the brain cells,
by using spectral data which is corrupted by Rician noise. In this work we
parametrize these transition distributions
 with diffusion tensors of either 2nd or 4th order.

We follow the Bayesian paradigm, choosing
  improper non-informative priors for tensors and noise parameters. Indeed,
in the Bayesian regularization of the tensor field  only very  little assumptions are needed,
namely an improper and isotropic Gaussian Markov random field prior, where the regularization parameters with scale invariant priors
are also estimated from the data.
This is very much in the spirit of E.T. Jaynes  
who advocated for Bayesian inference using objective priors, which  should be
 based on symmetries and on the maximum entropy principle when prior information is not available (\cite{jaynes}). 
 It is also not far from  the penalized maximum likelihood approach, with
the difference that we use as  Bayesian estimator the posterior expectations
rather than the Maximum A Posteriori configuration (which again could be obtained by
simulated annealing after adding a temperature parameter to our Gibbs-Metropolis algorithm).
 
Although  Bayesian regularization has already been  used in the diffusion-MRI 
literature, until now McMC  was not seen as 
a viable alternative for the analysis of  high  $b$-value diffusion-MR data.
To obtain diffusion  images, we need to process an huge  amount of data. Standard McMC
strategies like single site updates and random walk proposals were not  efficient  enough
to produce  whole brain images under the Rice noise model. 
By exploiting the properties of  Generalized Linear Models,
we are able to construct a Gaussian approximation to the full conditional distribution and
update simultaneously large blocks of tensor variables with high acceptance
rates. 
It is clear that our fully Bayesian approach, as well as all methods based on penalized maximum likelihood,
is computationally extensive compared with  multi-stage procedures where first the tensors
are estimated  independently, and only in a second step smoothing and interpolation
procedures are applied. However second-stage smoothing has  
its drawbacks, for example it depends on the choice of the tensor metrics, 
it can induce 
 unwanted effects as tensor swelling \citep{dryden}.
Nowadays there are   affordable options for acceleration, e.g.  adopting parallel computation on
a large computer cluster, 
 and computing with Graphical Processor Unit (GPU) \citep{GPU}. On the other hand, 
the acquisition of 
 MR-diffusion  data is very costly and  we cannot
keep a subject for hours inside the scanner, in order to get the most out of the  data
it makes sense to use more computational resources and perform an accurate Bayesian computation  under the exact noise model
combining estimation and adaptive regularization in single procedure.

We are currently working to extend our framework in several directions.
In a forthcoming paper, we have implemented the variational Bayes (VB) 
 approximation of the posterior distribution under the very same Bayesian hierarchical model discussed in this work.
We are also working on  positive definite tensor models, as the ternary quartic approach   
\citep{barmpoutis},\citep{ghosh_thesis},\citep{ghosh}, and on 
spherical harmonic expansions with variable dimensions, with  random  truncation levels  
at each voxel, and using  reversible-jump McMC to sample
 from the posterior \citep{green}. This would produce a  brain segmentation with classification of the voxels 
according to the tensors order.

\if
In this paper, we introduce the exact noise of diffusion MR data by a novel Bayesian model to estimate diffusion tensors of different order. 
An McMC Gibbs-Metropolis algorithm has been implemented for posterior computation. 
The model construction is based on an inverse statistical sampling idea 
from the diffusion MRI data, where the model \emph{always} stems from the (real) \emph{Rice distribution} of 
the MR measurements rather than other competitive, for which 
only partial measurements fit the models for tensor estimation. Our work provides a path to gain tensor information 
also from high-amplitude data corresponding to the signal to noise (SNR) and zero measurements. 
This contribution should serve as a starting point to understand detailed brain structures by diffusion tensor images.  
\nolinebreak
 
We also introduce a stochastic interpretation into the model for tensor estimation. 
Regularization transfuses two advantages into the model by considering the dependence between a voxel and its neighbours,
 and between a block of voxels and its neighbouring blocks. 1) Some tensors from the artefact/ corrupt measurements can be detected by the structures of their neighbours  2) Under fitted and/or over fitted estimates represented by abnormal tensor shapes (outliers) can be smoothed by their neighbours. 
Moreover, high-order Cartesian tensor modeling for higher order tensor estimation can be also used in our work.
\nolinebreak
These improvements provide a possible way to obtain more accurate tensor estimates,
 and thus produce more plausible DTI relative images by using all kinds of limited 
and costly diffusion MRI measurements. We believe our results will be a benefit also in clinical studies.

We also mention a few open questions for future study. 
\nolinebreak
One is high angular resolution of tensors do capture 
complex tissue structures in the brain. However, it should be noticed that when we fit a higher angular
 resolution diffusion imaging (HARDI) model for
to a voxel were the diffusion is monodirectional,   the precision of the  estimators is reduced,
 and high-order tensor modeling  increases the computation complexity. 
In the sense, some non-parametric method, such as spherical harmonics, can be considered. 
Furthermore, brain segmentation with classification of the voxels according t
o the order of the tensors may be a wiser proposal in tensor estimation. On the hand, 
modeling of characterization of diffusion tensors is also an interesting topic to fully understand water diffusion inside of brain.

\fi

\section*{Appendix} \label{appendix}
\appendix 
\numberwithin{equation}{section}
                                      
\section{Sampling from the reinforced Poisson distribution}\label{reiforced:sampling}

\begin{enumerate}

\item The standard way by using the cumulative distribution function:
\begin{align*}
X(\omega) = \min\biggl\{ n : \;\sum\limits_{k=0}^n  \frac{  \tau^{2k}  }{ (k! )^2} \; \ge  \; {}_0F_1( 1,\tau^2)  \; \omega \biggr\}.
\end{align*}
with $\omega$ uniformly distributed in $[0,1]$. This requires
evaluation of  the normalizing constant  ${}_0F_1( 1,\tau^2)$.

\item A direct but unefficient rejection method:

Generate $N \sim \mbox{Poisson}(\tau )$, accept it and set $X = N$
with probability $P_{\tau}(N'=N| N)=\exp(-\tau)\tau^N/N!$ 
where $N'$ is an independent copy of $N$, otherwise repeat until acceptance.

\item An improved rejection sampler, the one actually used.
Generate independenty $N\sim \mbox{Poisson}(\alpha )$ and $\omega$ uniform in $[0,1]$,\\
until 
\begin{align*}
     \frac{ \tau^{2N} }{ (N!)^2 }  \frac 1 { \pi_{\alpha}(N ) } = 
\frac{ (\tau^2 / \alpha )^N }{ N! } \exp(\alpha)\ge C(\alpha,\tau) \; \omega
\end{align*}

where
\begin{equation}
C(\alpha, \tau) := \max_n  \biggl\{ \exp( \alpha) \frac{ (\tau^2 / \alpha )^n }{ n! }  \biggr\}
= \frac{ ( \tau^2 /\alpha )^{n^*}  }{ n^* ! }  \exp( \alpha) 
\end{equation}
and $n^* = \lfloor \tau^2 / \alpha  \rfloor$  is the mode of a Poisson distribution with parameter
 $\tau^2/\alpha$, ( $\lfloor \cdot \rfloor$ denotes the floor function).
Return $X = N$.

For large $\tau$,  assuming apriori that
at optimality $\alpha \ll \tau^2$, by  using 
 Stirling's approximation $\log(n!)\approx ( n\log(n)-n)$,  we find
 that the proposal parameter $\alpha(\tau)=\tau$ is approximately optimal.
\end{enumerate}

\section*{Acknowledgements}

 We thank Professor Antti Penttinen and Salme K\"arkk\"ainen
for reviewing the manuscript, Touko Kaasalainen, Tarja Pohjasvaara, Veli-Pekka Poutanen and Oili Salonen
from the Radiology Unit  of Helsinki University Hospital for their invaluable collaboration in the data acquisition process.
We thank also 
J\"uri Lember and  Alexey Koloydenko for the interesting discussions.
The second author was funded  by the graduate school of Computations and Mathematical Science (COMAS) of the University of 
Jyväskylä. 
We are grateful to  the CSC-IT Center for Science Ltd. for the use of their
computer cluster, and  to the Finnish Doctoral Programme in Stochastics and Statistics (FDPSS) supporting
the project with travel grants.

\newpage

\begin{minipage}{\linewidth}
\centering
\captionof{table}{\label{bvalues}$b$-values and number of acquisitions.}
\begin{tabular}{|l|l |l| l| l|} \toprule[1.5pt]& \multicolumn{4}{|c|}{ Slice } 
\\ \hline
  $b$-value, $s/mm^2$     &   1  & 2  & 3  & 4 \\  \hline 
  0      &    3    & 3    & 3 & 2 \\ \hline
  62     &  3 $\times$ 32 & 3 $\times$ 32 &  3 $\times$ 32 & 2 $\times$ 32 \\ \hline
 249     & 3 $\times$ 32&  3 $\times$ 32 & 3 $\times$ 32 & 2 $\times$ 32 \\    \hline 
  560    & 3 $\times$ 32&  3 $\times$ 32 & 3 $\times$ 32 & 2 $\times$ 32 \\          \hline
   996   & 3 $\times$ 32&  3 $\times$ 32 & 3 $\times$ 32 & 2 $\times$ 32 \\   \hline 
   1556  & 3 $\times$ 32&  3 $\times$ 32 & 2 $\times$ 32 & 2 $\times$ 32 \\    \hline 
   2240  & 3 $\times$ 32&  3 $\times$ 32 & 2 $\times$ 32 & 2 $\times$ 32  \\     \hline 
 3049    & 3 $\times$ 32&  3 $\times$ 32 & 2 $\times$ 32 & 2 $\times$ 32 \\  \hline 
  3982   & 3 $\times$ 32&  3 $\times$ 32 & 2 $\times$ 32 & 2 $\times$ 32 \\     \hline 
   5040  & 3 $\times$ 32&  3 $\times$ 32 & 2 $\times$ 32 & 2 $\times$ 32 \\      \hline 
   6222  & 3 $\times$ 32&  3 $\times$ 32 & 2 $\times$ 32 & 2 $\times$ 32 \\    \hline 
   7529  & 3 $\times$ 32&  3 $\times$ 32 & 2 $\times$ 32 & 2 $\times$ 32 \\    \hline 
    8960 & 3 $\times$ 32&  3 $\times$ 32 & 2 $\times$ 32 & 2 $\times$ 32 \\      \hline 
  10516  & 3 $\times$ 32&  3 $\times$ 32 & 2 $\times$ 32 & 2 $\times$ 32 \\   \hline 
  12196  & 3 $\times$ 32&  3 $\times$ 32 & 2 $\times$ 32 & 2 $\times$ 32 \\   \hline 
   14000 & 3 $\times$ 32&  3 $\times$ 32 & 2 $\times$ 32 & 2 $\times$ 32 \\ \bottomrule[1.25pt]
        \end{tabular}
\bigskip

For each $b$-value and gradient direction we had 2-3  independent acquisitions, depending on the brain slice. 
\end{minipage}

\begin{minipage}{\linewidth}
 \centering 
\captionof{table}{\label{gradient:directions}Gradient directions}
\begin{tabular}{ |l |l |l| }\toprule[1.5pt]
     $u_x$ &$u_y$ & $u_z$  \\ \hline 
 -0.5000   & -0.5000  & -0.7071 \\    \hline 
   -0.5000 &  -0.5000   & 0.7071  \\  \hline 
    0.7071  & -0.7071  & -0.0000  \\  \hline 
   -0.6533  & -0.2706  & -0.7071 \\  \hline 
   -0.2087 &  -0.6756 &  -0.7071\\  \hline 
    0.0197 &  -0.7068 &  -0.7071\\  \hline 
    0.4212 &  -0.5679  & -0.7071\\  \hline 
    0.6899  & -0.1549  & -0.7071\\  \hline 
   -0.6535  & -0.2707  & -0.7069\\  \hline 
   -0.2929  & -0.7071 &  -0.6436\\  \hline 
    0.2945 &  -0.7064  & -0.6436\\  \hline 
    0.5150 &  -0.4861  & -0.7061\\  \hline 
    0.7071  & -0.2929 &  -0.6436\\  \hline 
   -0.7071  & -0.4725 &  -0.5261\\  \hline 
   -0.4725  & -0.7071 &  -0.5261\\  \hline 
    0.5555  & -0.6439  & -0.5261\\  \hline 
    0.7071   &-0.4725  & -0.5261\\  \hline 
   -0.7071 &  -0.7071  & -0.0002\\  \hline 
   -0.7071  & -0.4725 &   0.5261\\  \hline 
    0.7071   &-0.4725  &  0.5261\\  \hline 
    0.4725 &  -0.7071 &   0.5261\\  \hline 
   -0.7071  & -0.7071 &   0.0078\\  \hline 
   -0.6364 &  -0.4252 &   0.6436\\  \hline 
   -0.7060   &-0.7060  &  0.0547\\  \hline 
   -0.2929  & -0.7071 &   0.6436\\  \hline 
    0.2929  & -0.7071&    0.6436\\  \hline 
    0.7071  & -0.7071 &   0.0078\\  \hline 
    0.7071  & -0.2929  &  0.6436\\  \hline 
   -0.7063  & -0.7063  &  0.0489\\  \hline 
    0.0347  & -0.7063  &  0.7071\\  \hline 
    0.7071  & -0.7071  &  0.0115\\  \hline 
    0.7071  &  0.0000  &  0.7071\\
 \bottomrule[1.25pt]
\end{tabular}
\bigskip

 For each $b$-value, the  MR-signal was measured in these $32$ gradient directions.
\end{minipage}


\begin{thebibliography}{100}

\bibitem[Andersson J.L.R.(2008)]{andersson2008} Andersson J.L.R. 2008. 
Maximum a posteriori estimation of diffusion tensor parameters using a Rician noise model: Why how and but.
\textit{ Neuroimage } 42(4) 1340-1356.

\bibitem[Assemlal, H.E. et al. (2009)]{asselmal}
Assemlal, H.E., Tschumperl\'e D., Brun L. 2009. Efficient and robust computation of
PDF features from diffusion MR signal. \textit{ Medical Image Analysis} 13:5  715-729.

\bibitem[Barmpoutis A. et al. (2009)]{barmpoutis2}Barmpoutis, A. and Hwang, M.S. and Howland, D. and Forder, J.R. and Vemuri, B.C., 2009. 	
Regularized positive-definite fourth order tensor field estimation from DW-MRI.
\textit{ NeuroImage}, S153-S162. 

\bibitem[Barmpoutis A., Vemuri B.C. (2009)]{barmpoutis3} 
Barmpoutis A., Vemuri B.C., 2009. 	
Groupwise Registration and Atlas Construction of 4th-Order Tensor Fields Using the $\mathbb{R}^+$ Riemannian Metric.
\textit{
Medical Image Computing and Computer-Assisted Intervention – MICCAI 2009}, Springer Lecture Notes in Computer Science 5761, 640-647.  

\bibitem[Barmpoutis A., Vemuri B.C. (2010)]{barmpoutis} Barmpoutis A., Vemuri B.C., 2010.
A unified framework for estimating diffusion tensors of any order with symmetric positive-definite constraints.
\textit{ Biomedical Imaging: From Nano to Macro}  2010 IEEE Int. Symp, 1385-1388.

\bibitem[Basser PJ, Mattiello J, Le Bihan D.(1994)]{basser-94}  Basser P.J., Mattiello J., Le Bihan D., 1994.   Estimation of the effective self-diffusion tensor from 
the NMR spin echo. \textit{J. Magn. Reson. B} 103(3), 247-254. 

\bibitem[Basser P.J., Pajevic S. (2007) ]{basser-pajevic07}
Basser P.J., Pajevic S., 2007. 
Spectral decomposition of a 4th-order covariance tensor: Applications to diffusion tensor MRI.
\textit{ Signal Processing} 87, 220-236
  
\bibitem[Basser P.J., Pajevic S. (2003)]{basser-pajevic03}
Basser P.J., Pajevic S., 2003.
A normal distribution for tensor-valued random
variables: applications to diffusion tensor MRI.  
\textit{ IEEE Trans. Med. Imag.} 22 (7), 785-794.  

\bibitem[Behrens. T.E.J. et al. (2003)]{behrens2003} Behrens T.E.J. ,Woolrich M.W. ,Jenkinson M. ,
  Johansen-Berg H., Nunes R.G.,
Clare S. ,Matthews P.M. , Brady J.M., Smith S.M. 2003.  
Characterization and Propagation of Uncertainty in
Diffusion-Weighted MR Imaging.
\textit{Magnetic Resonance in Medicine }50:1077–1088.

\bibitem[Besag J. et al. (1991)]{besag_york_moille}
 Besag, J., York, J.,  Molli\'e A. (1991). 
Bayesian image restoration, with two applications in spatial statistics.
 {\it Ann. Inst. Statist. Math.} 43 (1) 1–59.

\bibitem[Besag J. et al. (1995)]{besag}
Besag J., Green P., Higdon D., Mengersen 
K. 1995. 
Bayesian computation and stochastic systems (1995). \textit{Statistical Science} 10:1, 3–66.

\bibitem[Burdette, J.H. et al. (2001)]{burdette} Burdette, J.H., Durden, D.D., Elster, A.D., Yen, Y.F., 2001. 	
High b-value diffusion-weighted MRI of normal brain.
\textit{J. Comput. Assi. Tomogr.} 25(4), 515.

\bibitem[Carr H.Y., Purcell E.M. (1954)]{carr} Carr H.Y., Purcell E.M., 1954. Effects of diffusion on free precession in nuclear magnetic 
resonance experiments. \textit{Phys. Rev.} 94(3), 630-638. 

\bibitem[Dryden I.L. et al. (2009)]{dryden}
Dryden I.L.,Koloydenko A., Zhou D. 2009.
Non-Euclidean statistics for covariance matrices, with applications to diffusion tensor imaging
\textit{Annals of Applied Statistics} 3 (3) 881-1231.

\bibitem[Frandsen, J. et al. (2007)]{FHLVV}
Frandsen, J., Hobolth, A., {\O}stergaard, L., Vestergaard-Poulsen, P., Jensen, E.B.V., 2007.
Bayesian regularization of diffusion tensor images.
\textit{Biostatistics} 8 (4), 784-799.

\bibitem[Geman S. and Geman D.(1984)]{geman} Geman S., Geman D., 1984. Stochastic Relaxation, Gibbs Distributions, and
the Bayesian Restoration of Images. \textit{IEEE Trans. Pattern Anal. Mach. Intell. } 6, 721-741.

\bibitem[Ghosh A. et al. (2009)]{ghosh} 
Ghosh A., Deriche R., Moakher M., 2009. 	
Ternary quartic approach for positive 4th order diffusion tensors revisited. \textit{
Biomedical Imaging: From Nano to Macro}, ISBI'09. IEEE Int. Sym., pp. 618-621.


\bibitem[Ghosh A (2011)]{ghosh_thesis} Ghosh A., 2011.\textit{ High Order Models in Diffusion MRI
and Applications.} PhD Thesis, Inria Sophia Antipolis.


\bibitem[Ghosh A. et al. (2012)]{GPD-inria}
Ghosh A., Papadopoulo T. and Deriche R., 2012.
Generalized Invariants of a 4th order tensor:
Building blocks for new biomarkers in dMRI.
\textit{Computational Diffusion MRI Workshop (CDMRI), MICCAI } 165-173.


\bibitem[Gradshteyn, I.S., Ryzhik, I.M. (2007)]{tables} 
Gradshteyn, I.S., Ryzhik, I.M., 2007.
\textit{Table of Integrals, Series, and Products, seventh edition.}
edited by Jeffrey, A., Zwillinger, D.  
Academic Press, pp. 918-920. 

\bibitem[Green P.J. (1995)]{green} Green P.J. 1995.
Reversible Jump Markov Chain Monte Carlo Computation and Bayesian Model Determination.  \textit{Biometrika}, Vol. 82, No. 4. 711-732.


\bibitem[Gudbjartsson H., Patz S. (2005)]{gudbjartsson} Gudbjartsson H., Patz S., 2005. 	
The Rician distribution of noisy MRI data.
\textit{Magn. Reson. Med.}, 34(6), 910-914.


 \bibitem[Hahn E.(1950)]{hahn} Hahn E., 1950. Spin echoes.\textit{ Phys. Rev.} 80, 580–594.

\bibitem[Hagmann P. et al. (2006)]{hagmann} 
Hagmann, P., Jonasson, L., Maeder, P., Thiran, J.P., Wedeen, V.J., Meuli, R., 2006.
Understanding Diffusion MR Imaging Techniques: From Scalar Diffusion-weighted Imaging to Diffusion Tensor Imaging and Beyond,
\textit{Radiographics,} 26(suppl 1), S205-S223.


\bibitem[Hastings W.K. (1970)]{hastings}
Hastings W.K., 1970.
Monte Carlo sampling methods using Markov chains and their applications. 
\textit{Biometrika} 57 (1), 97-109.


\bibitem[Henkelman R.M. (1985)]{henkelman} Henkelman R.M., 1985. 	
Measurement of signal intensities in the presence of noise in MR images.
\textit{Med. phys.,} 12(2), 232-233.



\bibitem[Hern\'andez M. et al. (2013) ]{GPU}
Hern\'andez M.,
  Guerrero G.D.,
Cecilia      J.M.,
 Garcia J.M.,
Inuggi A.,
 Jbabdi S.,
 Behrens T.E.J. 2013.
Accelerating Fibre Orientation Estimation from 
Diffusion Weighted Magnetic Resonance Imaging Using GPUs.
\textit{PlosOne} 0061892.



\bibitem[Huisman, T.A.G.M. et al.(2006)]{huisman} 	
Huisman, T.A.G.M., Loenneker, T., Barta, G., Bellemann, M.E., Hennig, J., Fischer, J.E., Il’yasov, K.A., 2006.
Quantitative diffusion tensor MR imaging of the brain: field strength related variance of apparent diffusion coefficient (ADC) a
nd fractional anisotropy (FA) scalars.\textit{
Eur. radiol.} 16(8), 1651-1658.


\bibitem[Jaynes E.T. (2002)]{jaynes} Jaynes E.T., 2002. \textit{ Probability, the Logic of Science}. Cambridge University Press.



\bibitem[Jeffreys H. (1961)]{jeffreys} 
Jeffreys H., 1961. \textit{
Cartesian Tensors.}
Cambridge University Press.

\bibitem[Jian B., Vemuri B.C. (2007)]{Jian-vemuri} 
Jian B., Vemuri B.C., 2007.
Multi-fiber Reconstruction from Diffusion MRI Using Mixture of Wisharts and Sparse Deconvolution.\textit{
IPMI}, pp. 384-395.

\bibitem[Jones D.K., Basser P.J. (2004)]{jones} Jones D.K., Basser P.J., 2004. 	
"Squashing peanuts and smashing pumpkins": How noise distorts diffusion-weighted MR data.
\textit{Magn. Reson. Med.} 52(5), 979-993.

\bibitem[Krissian K. Aja-Fernandez S. (2009)]{krissian}
Krissian, K., Aja-Fernández S. (2009). Noise-driven anisotropic diffusion filtering of MRI. 
\textit{IEEE Transactions on Image Processing}  18  (10)  2265-2274.

\bibitem[Kaipio J., Somersalo E. (2005)]{kaipio_somersalo}
Kaipio J., Somersalo E., 2005. \textit{
Statistical and Computational Inverse Problems}. Springer.

\bibitem[Koay et al. (2006)]{koay}   Koay C.G., Chang L.C., Carew J.D., Pierpaoli C., Basser P.J.,   2006.
A unifying theoretical and algorithmic framework for least squares methods of estimation in diffusion tensor imaging. \textit{
Journal of Magnetic Resonance } 182(1): 115-125.

\bibitem[Koay C.G., \"Ozarslan E., Basser P.J. (2009)]{peter} Koay C.G., \"Ozarslan E., Basser P.J., 2009. 	
A signal transformational framework for breaking the noise floor and its applications in MRI,
\textit{J. Magn. Reson.}, 197(2), 108-119.

\bibitem[Landman B. et al (2007)]{landman} Landman B. 
  Bazin P-L. and Prince J. 2007.
Diffusion tensor estimation by maximizing Rician likelihood. 
\textit{11th IEEE  International Conference on Computer Vision  ICCV 2007}.

\bibitem[Lange K. (2013)]{lange} Lange K (2013). \textit{ Optimization } (2nd Edition). Springer Texts in Statistics Vol. 95. Springer New York.

\bibitem[Lauterbur P.C. (1973)]{Lauterbur}
Lauterbur P.C., 1973. 
Image formation by induced local interactions: examples employing nuclear magnetic resonance. 
 \textit{ Nature} 242 (5394), 190-191.

\bibitem[Lauwers L. et al. (2010)]{Lauwers}  Lauwers L., Barb\'e K., Van Moer W., Pintelon R., 2010.
    Analyzing Rice distributed functional magnetic resonance imaging data: a Bayesian approach.
  \textit{ Measurement Science \& Technology } 21, 115804.

\bibitem[Le Bihan D et al. (1986)]{lebihan86}
Le Bihan D., Breton E., Lallemand D., Grenier P., Cabanis E., Laval-Jeantet M., 1986.
MR imaging of intravoxel incoherent motions: application to diffusion 
and perfusion in neurologic disorders. \textit{ Radiology } 161(2), 401-407.

\bibitem[Leemans A et al.  (2009)]{Alexander}
Leemans, A., Jeurissen, B., Sijbers, J., Jones, D.K., 2009. 
ExploreDTI: a graphical toolbox for processing, analyzing, and visualizing diffusion MR data.
 \textit{ Proc. Intl Soc. Mag. Reson. Med } 3537 Hawaii, USA. 

\bibitem[Liao Y. et al. (2010)]{Liao}
Liao, Y., Tang, J., Ma, M., Wu, Z., Yang, M., Wang, X., Liu, T., Chen, X., Fletcher, P.C., Hao, W., 2010.
Frontal white matter abnormalities following chronic ketamine use: a diffusion tensor imaging study.
\textit{ Brain}, 133(7), 2115-2122.

\bibitem[Marinucci D., Peccati G. (2011)]{peccati:marinucci}
Marinucci D, Peccati G.  2011.
\textit{ Random fields on the sphere. Representation,
 limit theorems and cosmological applications}.
 London Mathematical Society Lecture Note Series, 389. Cambridge University Press, Cambridge, 


\bibitem[McCullagh, P., Nelder, J.A. (1989)]{mc_cullagh-nelder}
McCullagh, P., Nelder, J.A., 1989.  \textit{ 
Generalized linear models} 2nd Edition. 
Chapman \& Hall/CRC.

\bibitem[Metropolis N. et al. (1953)]{metropolis}
Metropolis N., Rosenbluth A.W., Rosenbluth M.N., Teller A.H. and Teller E., 1953. 
Equation of State Calculations by Fast Computing Machines. \textit{ 
J.Chem. Phys.} 21 (6), 1087-1092.

\bibitem[Moakher M. (2009)]{moakher} Moakher M., 2009. 	
The algebra of fourth-order tensors with application to diffusion MRI.
In: Laidlaw D.H., Weickert J. (Eds.) \textit{ 
Vis. Process. Tensor Fields Advan. Perspect.}, Springer, pp. 57.

\bibitem[Mori S., Tournier J.D. (2014)]{Mori}Mori S., Tournier J.D., 2014. \textit{ 
Introduction to Diffusion Tensor Imaging and Higher Order Models}, 2nd Edition.
Elsevier Science. 


\bibitem[Moseley ME et al. (1990)]{moseley} Moseley M.E., Cohen Y., Kucharczyk J., Mintorovitch J., Asgari H.S., Wendland M.F., 
Tsuruda J., Norman D., 1990. Diffusion-weighted MR imaging of anisotropic water diffusion in 
cat central nervous system.\textit{  Radiology } 176(2), 439-445.


\bibitem[Nummelin E. (2002)]{nummelin}
Nummelin E., 2002. 
MC for MCMC'ists. \textit{ Int. Stat. Rev.} 70 (2), 215-240.

\bibitem[\"Ozarslan E., Mareci T.H. (2003)]{ozars} \"Ozarslan E., Mareci T.H., 2003. 	
Generalized diffusion tensor imaging and analytical relationships between diffusion tensor imaging and high angular resolution diffusion imaging.\textit{ 
Magn. Reson. Med.} 50 (5), 955-965.

\bibitem[\"Ozarslan E. et al. (2005)]{evren} \"Ozarslan E., Vemuri C. B., and Mareci T.H., 2005. 	
Generalized scalar measures for diffusion MRI using trace, variance, and entropy.\textit{ 
Magn. Reson. Med. } 53 (4), 866--876.

\bibitem[Pajevic S., Basser P.J. (2003)]{pajevic} Pajevic S., Basser P.J., 2003. 	
Parametric and non-parametric statistical analysis of DT-MRI data. \textit{ 
J. Magn. Reson. } 161 (1), 1-14.


\bibitem[Qi L. et al. (2010)]{qi} Qi L. Yu G., Wu E.X., 2010. 	
Higher order positive semidefinite diffusion tensor imaging. \textit{ 
SIAM J.Imag. Sci.}, 3(3), 416-433.


\bibitem[Robert, C.P., Casella G. (2004)]{robert-casella} Robert, C.P., Casella G. 2004.
\textit{Monte Carlo Statistical Methods, 2nd Edition}, Springer Text in Statistics.

\bibitem[Rue, H., Held, L. (2005)]{rue} 
Rue, H., Held, L., 2005.\textit{ 
Gaussian Markov Random Fields: Theory and Applications}. Chapman \& Hall/CRC, Boca Raton, FL.

\bibitem[Salvador, R. et al. (2004)]{salvador} 
Salvador, R., Pena, A., Menon, D.K., Carpenter, T.A., Pickard, J.D., Bullmore, E.T., 2004.
Formal characterization and extension of the linearized diffusion tensor model.\textit{ Hum. brain mapp.}, 24(3), 144-155.



\bibitem[Spiegelhalter, D.J. et al.(2002)]{spiegelhalter}
 Spiegelhalter, D.J.,  Best, N.G., Carlin, B.P.  van der Linde, A., 2002. 
Bayesian measures of model complexity and fit. \textit{ Journal of the Royal Statistical Society, B } 64 (4): 583–639.


\bibitem[Stejskal E.O., Tanner J.E. (1965)]{stejskal}
Stejskal E.O., Tanner J.E., 1965. Spin diffusion measurements: spin echoes in the presence of 
time-dependent field gradient. \textit{ J. Chem. Phys.}, 42(1), 288-292.


\bibitem[Torrey H. (1956)]{torrey} Torrey H., 1956. Bloch equations with diffusion terms.\textit{ Phys. Rev.} 104,563–565.

 
\bibitem[Veraart, J. et al. (2011)]{Jelle}Veraart, J., Van Hecke, W., Sijbers, J., 2011. 	
Constrained maximum likelihood estimation of the diffusion kurtosis tensor using a Rician noise model.\textit{ Magn. Reson. Med.}, 66(3), 678-686.

\bibitem[Zhu H. et al. (2007)]{ibrahim} Zhu H., Zhang H., Ibrahim J.G., Peterson B.S., 2007. 	
Statistical analysis of diffusion tensors in diffusion-weighted Magnetic resonance imaging Data.\textit{ JASA } 102 (480), 1085-1102.  
  
\end{thebibliography}
 \end{document}